%% file: main.tex
\documentclass{article}
\pdfoutput=1
\usepackage{lineno,hyperref}
\usepackage{authblk}
\usepackage{geometry}
\newgeometry{vmargin={25mm}, hmargin={28mm,28mm}}

\usepackage{physics}
\usepackage{xcolor}
\usepackage{amsthm, amsfonts}
\usepackage{mathrsfs, mathtools}

\usepackage{graphicx}
\usepackage{multirow} 
\usepackage{relsize} 
\usepackage{caption, subcaption} 
\usepackage{siunitx}
\usepackage{cleveref}

\title{\huge{A deterministic adjoint-based semi-analytical algorithm for fast
       response change computations in proton therapy}}

\author[1,2]{Tiberiu Burlacu}
\author[1,2]{Danny Lathouwers\footnote{Both authors contributed equally}}
\author[1,2]{Zolt\'{a}n Perk\'{o}\protect\footnotemark[1]}

\affil[1]{\footnotesize{Delft University of Technology, Faculty of Applied Sciences, Delft, The Netherlands}}
\affil[2]{\footnotesize{HollandPTC consortium\footnote{HollandPTC consortium – Erasmus Medical Center, Rotterdam, Holland Proton Therapy Centre, Delft, Leiden University Medical Center (LUMC), Leiden and Delft University of Technology, Delft, The Netherlands}, Delft, The Netherlands}}

\date{\small{\today}}

\begin{document}

\maketitle

\begin{abstract}
In this paper we propose a solution to the need for a fast particle transport algorithm in Online Adaptive Proton
Therapy capable of cheaply, but accurately computing the changes in patient dose metrics as a result of changes in the system parameters. We obtain the proton phase-space density through the product of the
numerical solution to the one-dimensional Fokker-Planck equation and the
analytical solution to the Fermi-Eyges equation. Moreover,
a corresponding adjoint system was derived and solved for the adjoint flux.
The proton phase-space density together with the adjoint flux and
the metric (chosen as the energy deposited by the beam in a variable region of
interest) allowed assessing the accuracy of our algorithm to different
perturbation ranges in the system parameters and regions of interest.
The algorithm achieved negligible errors
($\qtyrange{1.1E-6}{3.6E-3}{\percent}$) for small perturbation ranges
($\qtyrange{-40}{+40}{HU}$) and small to moderate errors
($\qtyrange{3}{17}{\percent}$) -- in line with the well-known limitation of adjoint approaches -- for large perturbation ranges
$(\qtyrange{-400}{+400}{HU})$ in the case of most clinical interest where the
region of interest surrounds the Bragg peak.
Given these results coupled with the capability of further improving the timing
performance it can be concluded that our algorithm presents a viable solution
for the specific purpose of Online Adaptive Proton Therapy.
\end{abstract}

\include{./src/introduction}

\include{./src/theory_approximating_the_lbe}

\include{./src/theory_fokker_planck}

\include{./src/theory_fermi_eyges}

\include{./src/theory_adjoint}

\include{./src/results}

\include{./src/conclusion}

\appendix

\bibliographystyle{elsarticle-num}
\bibliography{bib/bibliography}

\end{document}

%% file: src/introduction.tex
\section{Introduction}

\subsection{Charged particle transport}

The importance of studying charged particle transport
is perhaps best illustrated by its applications in a wide-ranging set of
fields such as radiation protection, radiotherapy, space radiation
shielding, electron and ion beam microscopy, or surface analysis and lithography
\cite{zheng-mingOverviewTransportTheory1993a}. The goal of charged particle
transport problems is to obtain the phase-space density of particles using
modelled or empirically sourced reaction cross-sections. The phase-space
density of particles provides a complete description of the particle fluence
and all quantities of interest that can be derived from it. To obtain a
general integro-differential equation that describes
the phase-space density of particles in a scattering medium the
collision-free Boltzmann equation \cite{duderstadtTransportTheory1979} is
altered to account for collisions via a scatter term. This equation is
Boltzmann's general transport equation and for most realistic applications its
solution is highly complex.

In practice several application-dependent approximations are applied
to the Boltzmann equation in order to obtain an analytical or numerical
solution, with each approaches having their individual trade-offs. For example,
while Monte Carlo (MC) methods have as
advantages high precision and an ease of understanding, their main
disadvantage is the slow computation times which deem them inapplicable in
many scenarios \cite{zheng-mingOverviewTransportTheory1993a}, especially when
(near) real-time calculations are necessary
\cite{botasOnlineAdaptionApproaches2018}. Diametrically opposite to MC methods
from a computational expense standpoint are analytical methods such as the
pencil beam approaches, which are obtained through
fits and approximations. As expected, these methods trade-off the high precision
for the low computational expense. In between these two extremes lie several
numerical or semi-numerical approaches, such as the moment method or the
phase space time evolution method \cite{cordaroApplicationPhaseSpace1972}.

The focus of this paper is on a combination
of numerical and analytical methods (the pencil beam and energy straggling
methods) that are deemed promising for fulfilling the needs of our specific
application, so called online adaptive proton therapy, which is currently the
state-of-the-art form of radiotherapy.


\subsection{Particle Transport Needs In Online Adaptive Proton Therapy}

Within the field of radiation therapy, proton therapy (PT) has emerged as an
alternative to conventional photon radiotherapy for cancer treatment due to its
promises of increased dose conformity and lowered doses achievable in healthy
tissues \cite{paganettiProtonTherapyPhysics2016}. These benefits are due to the
presence of the Bragg peak (BP) in the depth-dose distribution, as charged
particles deposit most of their energy within a small volume near the end of
their range. The Bragg peak however also makes proton doses highly susceptible
to uncertainties
\cite{lomaxIntensityModulatedProton2008,perkoFastAccurateSensitivity2016}.
Some of the common sources of
range uncertainties are related to CT imaging,
treatment delivery or changes in the anatomy of the patient
\cite{paganettiRangeUncertaintiesProton2012}.

Currently, the state-of-the-art in
dealing with uncertainties in clinical practice is to apply robust optimization
\cite{rojo-santiagoAccurateAssessmentDutch2021,
voortRobustnessRecipesMinimax2016}. In robust
optimization irradiation plans are optimized such that they ensure good
performance of the plan under even the most extreme uncertainty scenarios
\cite{unkelbachRobustProtonTreatment2018}. Due to the complexity of the
potential scenarios however, certain scenarios -- such as anatomical variations
(e.g., weight loss over the course of typically weeks long treatments) -- are
typically not accounted for \cite{paganettiAdaptiveProtonTherapy2021}. Most
importantly, robust planning essentially enlarges the high dose volume around
the tumor, increasing the dose in the surrounding healthy tissues, which in
turn increases the probability of detrimental side effects
\cite{vandewaterPriceRobustnessImpact2016a}.

The ideal solution would be to use Online Adaptive Proton Therapy (OAPT)
instead. In OAPT, a daily CT scan of the patient is acquired and within 30
seconds (the time for a robotic arm to move the patient from the in-room CT
scanner to the irradiation location) a new, fully re-optimized plan is created
\cite{botasOnlineAdaptionApproaches2018}. Having up-to-date anatomical
information allows accurately targeting the tumor
\cite{paganettiAdaptiveProtonTherapy2021} without needing robust planning,
leading to smaller irradiated volumes and fewer side effects. Unfortunately
however, the computational expense of dose calculations and plan
re-optimization \cite{menGPUbasedUltrafastDirect2010}, and the time needed for
the presently mostly manual plan quality assurance (QA)
\cite{barrettPracticalRadiotherapyPlanning2009} is far larger than 30 seconds,
making such workflows currently clinically infeasible.

Fast proton transport methods that are accurate in highly heterogeneous patient geometries are key to overcome these computational and QA related bottlenecks, and represent one of the missing enabling technology for online adaptive workflows and further improving cancer treatments. First, they are necessary for re-optimization, as plan optimization requires the dose distribution from each of the typically hundreds or even thousands of individual proton beams as input \cite{schwarzTreatmentPlanningProton2011}. Second, they are crucial for replacing the current manual, measurement based plan QA with fast computational alternatives. Traditional plan QA measurements assess the differences between planned and delivered doses in order to ensure they are within the clinically acceptable $\pm \SI{3}{\percent}$ \cite{gottschalkPassiveBeamSpreading2004} range and that the irradiation delivery system functions as intended \cite{frankProtonTherapyIndications2020}. Since manual measurements are clearly infeasible in OAPT, independent dose calculation methods \cite{liUseTreatmentLog2013} have been proposed as a viable alternative, showing similar precision when using accurate MC transport methods \cite{meierIndependentDoseCalculations2015b}. As further advantage, such automated QA procedures yield clinically more relevant metrics than measurements and could potentially even increase clinical throughput and treatment accessibility \cite{meijersFeasibilityPatientSpecific2020}. While the benefits of automated QA procedures based on independent dose calculation and machine log-files (measured the outgoing radiation from the treatment machine) are clear, MC calculations \cite{matterAlternativesPatientSpecific2018a}, even when multi-threaded \cite{meijersFeasibilityPatientSpecific2020} are not fast enough to perform (near) real-time QA necessary in the OAPT workflow.

\subsection{A Semi-Analytical Adjoint-Based Deterministic Algorithm For OAPT}

To overcome these issues we propose a semi-analytical
adjoint-based deterministic algorithm that could serve as (near) real-time plan QA, using machine log-files and the patient geometry. The semi-analytical component aims to provide a balance between the accuracy of MC algorithms and the speed of analytical dose calculation algorithms. The adjoint component aims to provide real-time quality assurance through efficient computations of the effect of perturbations in the system parameters (beam spatial and energy spread, its particle number or the
patient geometry) on the desired clinical metrics (dose, or more complex responses).

The semi-analytical component has as a starting point, similarly to the MC
algorithms, the Linear Boltzmann Equation (LBE). Through the continuous
slowing down, energy straggling and Fokker-Planck approximations
the LBE can be reduced to two partial differential equations (PDEs).
One of the PDEs is the one-dimensional Fokker-Planck (FP) equation
while the other one is the Fermi-Eyges (FE) equation. The advantage of this
approach is threefold. First, the approach derives a system which is described
by two PDEs. The presence of the PDEs (as opposed to for example a machine
learning (ML) based dose engine \cite{pastor-serranoMillisecondSpeedDeep2022}) allows the application adjoint methods.
Using functional analysis an adjoint
system can be derived which can be used to avoid the expensive process
of re-computing the solution to the two PDEs for each new set of
system parameters. Second, the physical approximation will not
suffer from the typical drawbacks of ML models such as out-of-distribution
samples i.e. will remain accurate despite the input not being already seen
by the algorithm. Third, while the one-dimensional FP equation requires
a numerical solution the FE equation has a known analytical solution
\cite{eygesMultipleScatteringEnergy1948, brahmeSimpleRelationsPenetration1975}.
This coupling will ensure the computational effectiveness as the lateral
part of the proton flux is computed through a straightforward function
evaluation.

\subsection{Paper outline}

Section \ref{sec:system_model} covers the theoretical background of reducing
the LBE to two simplified PDEs. Section \ref{sec:1DFP} describes the
one-dimensional Fokker-Planck equation and its numerical solution while Section
\ref{sec:FE} covers the Fermi-Eyges solution. In Section \ref{sec:response}
the application of the functional analysis framework
for the derivation of the adjoint system with its associated adjoint solution
is detailed, the solution methodology of the adjoint system is explained
and the response change computations due to perturbations in
the system parameters are given. Section \ref{sec:results_discussion} covers
benchmarks of our own algorithm versus TOPAS and Bortfeld's
algorithm and provides comparisons between the forward and adjoint
computation of the response changes due to system parameter perturbations.
Lastly, Section \ref{sec:conclusion} contains some conclusions and future
intended research directions.

%% file: src/theory_approximating_the_lbe.tex
\section{The system model}
\label{sec:system_model}
The physical system under consideration is given by a proton beam irradiating
the patient. This system can be characterized through the (steady-state) LBE,
the validity of which for PT has been discussed by Borgers
\cite{borgersRadiationTherapyPlanning1999}.
The LBE describes the proton balance in an
arbitrary volume. Its derivation is obtained by
equating all the gain and loss mechanisms for protons with a certain energy
$E$ in $\dd E$ and direction $\vu*{\Omega}$ in $\dd \vu*{\Omega}$ in an
arbitrary volume $V$ with a boundary denoted by $\partial V$ as outlined
by Duderstadt \& Hamilton \cite{duderstadtNuclearReactorAnalysis1991a}.
The equation is an integro-differential equation for the proton flux
($\varphi = v n$) with $v$ the proton speed and $n(\vb{r}, E, \vu{\Omega}, t)$
the angular proton density,
\begin{align}
\label{eq:LBE}
	\vu*{\Omega} \cdot \grad{\varphi}
	+ \Sigma_t(\vb{r},E) \varphi(\vb{r},E,\vu*{\Omega})
	&=
	\int\limits_{4\pi} \dd \vu*{\Omega}' \int\limits_0^\infty \dd E'
	\Sigma_s(E' \rightarrow E, \vu*{\Omega}' \rightarrow \vu*{\Omega})
	\varphi(\vb{r},E',\vu*{\Omega}') + s(\vb{r},E,\vu*{\Omega}) \\
	\text{BC: } \varphi(\vb{r}_s,E,\vu*{\Omega},t) &= 0
	\text{ if } \vu*{\Omega} \cdot \vu{e}_s < 0 \text{ with } \vb{r}_s
	\in \partial V,
\end{align}
where $ \Sigma_t $ is the total macroscopic cross section, $ \Sigma_s $ is the
macroscopic double differential scattering cross section and $ s $ is the
source of protons.

Currently, the LBE in its form is computationally expensive to solve.
A first step is to divide the total $ \Sigma_t $ and scatter $ \Sigma_s $
cross sections according to the main interactions that a proton undergoes
as it propagates through the medium, namely $ \Sigma_t = \Sigma_{a} +
\Sigma_{e} + \Sigma_{in}$ where $ \Sigma_{a} $ is the catastrophic (absorption)
scatter cross section, $ \Sigma_{e} $ is the elastic scatter cross section
between the incident protons and the nuclei of tissue, $ \Sigma_{in} $ is
the inelastic scatter cross section between the incident protons and
atomic electrons. By doing so, Equation \ref{eq:LBE} can be written as
\begin{align}
	\vu*{\Omega} \cdot \grad{\varphi} &=
	\int\limits_{4\pi} \dd \vu*{\Omega}' \int\limits_E^\infty \dd E'
	\Sigma_a(E' \rightarrow E, \vu*{\Omega}' \rightarrow \vu*{\Omega})
	\varphi(\vb{r},E',\vu*{\Omega}')
	- \Sigma_a(\vb*{r}, E)\varphi(\vb{r},E',\vu*{\Omega}') \nonumber \\
	&+ \int\limits_{4\pi} \dd \vu*{\Omega}'
	\Sigma_e(\vb{r}, E, \vu*{\Omega}' \rightarrow \vu*{\Omega})
	\varphi(\vb{r},E,\vu*{\Omega}')
	- \Sigma_e(\vb*{r}, E) \varphi(\vb{r},E,\vu*{\Omega}) \label{eq:LBE_split} \\
	&+ \int\limits_0^\infty \dd E'
	\Sigma_{in}(\vb*{r}, E+ Q \rightarrow E, \vu*{\Omega})
	\varphi(\vb{r},E + Q,\vu*{\Omega})
	- \Sigma_{in}(\vb*{r}, E) \varphi(\vb{r},E,\vu*{\Omega}). \nonumber
\end{align}
In this splitting it is assumed that the energy transfer in
Coulomb elastic scatter interactions is negligible and that the angular
deflection in Coulomb inelastic scatter interactions is negligible
\cite{zheng-mingOverviewTransportTheory1993a}.

The next step is to apply approximations to each of the collision integrals
in Equation \ref{eq:LBE_split}. The inelastic scatter integral is approximated
using the Continuous Slowing Down Approximation (CSDA) and the Energy-loss
Straggling (ELS) approximation \cite{zheng-mingOverviewTransportTheory1993a}.
Therafter, the Fokker-Planck approximation is applied to the elastic scatter
angular integral and in the elastic scattering cross section $ \Sigma_e(E,
\vu*{\Omega} \cdot \vu*{\Omega}') $ the energy is replaced by the
depth-dependent mean energy $ E_a(z) $
\cite{gebackAnalyticalSolutionsPencilBeam2012a,
zheng-mingOverviewTransportTheory1993a}. The catastrophic scatter integral is
neglected completely with only the catastrophic scatter cross section absorption
term remaining. Applying these approximations to the LBE reduces the
integro-differential equation to the following PDE
\begin{align}
	\label{eq:LBE_PDE}
	\pdv{\varphi}{z} + \Omega_x \pdv{\varphi}{x} &+ \Omega_y \pdv{\varphi}{y}
	- \pdv{S(\vb*{r}, E)\varphi}{E}
	- \frac{1}{2} \pdv[2]{T(\vb*{r}, E) \varphi}{E}
	+ \Sigma_a(\vb*{r}, E) \varphi
	- \Sigma_{tr}(E_a(z)) \qty(\pdv[2]{\varphi}{\Omega_x}
	+ \pdv[2]{\varphi}{\Omega_y}) = 0,
\end{align}
where $S(\vb*{r}, E)$ is the stopping power, $T(\vb*{r}, E)$ is the
straggling coefficient, $\Sigma_a$ is the absorption cross section and
$\Sigma_{tr}$ is the transport cross section. The resulting PDE is linear in
the dependent variable $ \varphi $ which in turn depends on the six independent
system variables $ \vb{r},\vu*{\Omega}, E $.

We follow Geb\"{a}ck and Asadzadeh's and write
\cite{gebackAnalyticalSolutionsPencilBeam2012a}
\begin{align}
	\label{eq:flux_ansatz}
	\varphi = \varphi_{FE}(\vb*{r},\vu*{\Omega}) \cdot \varphi_{FP}(z,E).
\end{align}
Substitution in Equation \ref{eq:LBE_PDE} results in
\begin{align}
		\Upsilon\qty(\varphi_{FE}) \cdot \varphi_{FP}
	+ \varphi_{FE} \cdot \text{1DFP}(\varphi_{FP}) = 0,
\end{align}
where $ \Upsilon\qty(\varphi_{FE}) $ is the Fermi-Eyges equation and
$ \text{1DFP}(\varphi_{FP}) $ is the one-dimensional Fokker-Planck equation.
In order to avoid the trivial solution both of these equations must be set
to zero. Specifically,
\begin{align}
	\Upsilon\qty(\varphi_{FE}) = \pdv{\varphi_{FE}}{z}
	+ \Omega_x \pdv{\varphi_{FE}}{x} + \Omega_y \pdv{\varphi_{FE}}{y}
	- \Sigma_{tr}(E_a(z)) \qty(\pdv[2]{\varphi_{FE}}{\Omega_x}
	+ \pdv[2]{\varphi_{FE}}{\Omega_y}) = 0.
\end{align}
and
\begin{align}
	\text{1DFP}(\varphi_{FP}) = \pdv{\varphi_{FP}}{z}
	- \pdv{S(E) \varphi_{FP}}{E} - \frac{1}{2} \pdv[2]{T(E) \varphi_{FP}}{E}
	+ \Sigma_a(E) \varphi_{FP} = 0.
\end{align}

Using the solution of these two equations, the response of the system can be
defined which in this case was chosen as the energy deposited
in a certain region of interest (ROI). The method is applicable to other, more
general, responses (defined as functionals or operators) as long as the
chosen response satisfies a weak Lipschitz condition in the system state
vector and parameters \cite{cacuciSensitivityUncertaintyAnalysis2003a}.
In the case of this work, the response $R$ is given by
\begin{align}
 	R(\varphi) &= - \int\limits_{ROI} \dd V
								 \int\limits_{4 \pi} \dd \vu{\Omega}
								 \int\limits_{E_{min}}^{E_{max}} \dd E \qty [
								 \pdv{S(E)\varphi}{E} + \frac{1}{2} \pdv[2]{T(E) \varphi}{E}
								 - E \Sigma_a \varphi ] \nonumber \\
						 &= - \int\limits_{ROI} \dd V
						      \int\limits_{4 \pi} \dd \vu{\Omega} \varphi_{FE}
									\int\limits_{E_{min}}^{E_{max}} \dd E \qty [
									\pdv{S(E)\varphi_{FP}}{E}
									+ \frac{1}{2} \pdv[2]{T(E) \varphi_{FP}}{E}
 								 - E \Sigma_a \varphi_{FP} ],
\end{align}
where in the last equality Equation \ref{eq:flux_ansatz} was employed.

%% file: src/theory_fokker_planck.tex
\section{Approximating the one-dimensional Fokker-Planck equation}
\label{sec:1DFP}
The one-dimensional Fokker-Planck equation is a convection-diffusion
equation in energy whose character is well suited for Discontinuous
Galerkin (DG) methods. Consequently, its semi-discrete form was
obtained using the Symmetric Interior Penalty Galerkin (SIPG). The main
advantages of the SIPG method over other finite element methods (FEM)
are the relative ease with which the approximating polynomial
can be changed on different mesh elements, the fact that the method
allows unstructured or adaptive meshes, and that the method satisfies a
local energy balance (as opposed to the global energy balance satisfied
by continuous Galerkin methods) \cite{riviereDiscontinuousGalerkinMethods2008}.
The semi-discrete form was solved using the Crank-Nicholson (CN) method which
is a second order accurate implicit finite difference method. The advantage of
the CN method is that in a geometry that is piece-wise constant it relies on
only one of the previous points (as opposed to schemes such as the
Backward Differentiation Formula 2 that require two previous points
for the same order of accuracy \cite{suliIntroductionNumericalAnalysis}).

The one-dimensional Fokker-Planck equation can be written in a more standard
convection-diffusion form
\begin{align}
	\pdv{\varphi_{FP}}{z} - \pdv{S^*(E) \varphi_{FP}}{E}
	- \pdv{E} \qty(T^*(E) \pdv{\varphi_{FP}}{E})
	+ \Sigma_a(E) \varphi_{FP} = 0,
	\label{eq:1DFP}
\end{align}
where the modified stopping power $S^* = S + \frac{1}{2} \dv{T}{E} $ and the
modified straggling coefficient $T^* = T/2$ are introduced. To simplify
notation, the stars will from here on be dropped.
Moreover, it is Equation \ref{eq:1DFP} that will from now on be referred
to as the one-dimensional Fokker-Planck equation. This equation can also
be written in a short-hand form as
\begin{equation*}
	L(\vb*{\alpha}) \varphi = 0
\end{equation*}
where the vector of system parameters $\vb*{\alpha} \in \mathscr{L}_2$
and the differential operator acting on the flux $L$ are introduced as
\begin{align*}
	L(\vb*{\alpha})(\cdot)
	&= \pdv{z}(\cdot)
	- \pdv{S^* (\cdot)}{E}
	- \pdv{E} \qty[T^* \pdv{(\cdot)}{E}]
	+ \Sigma_a (\cdot),  \\
	\text{ and }
	\vb*{\alpha} &= \qty(S^*(E), T^*(E), \Sigma_a(E)).
\end{align*}

\subsection{Domain definition and discretization}
\begin{figure}[!h]
	\centering
	\includegraphics[width=.7\linewidth]{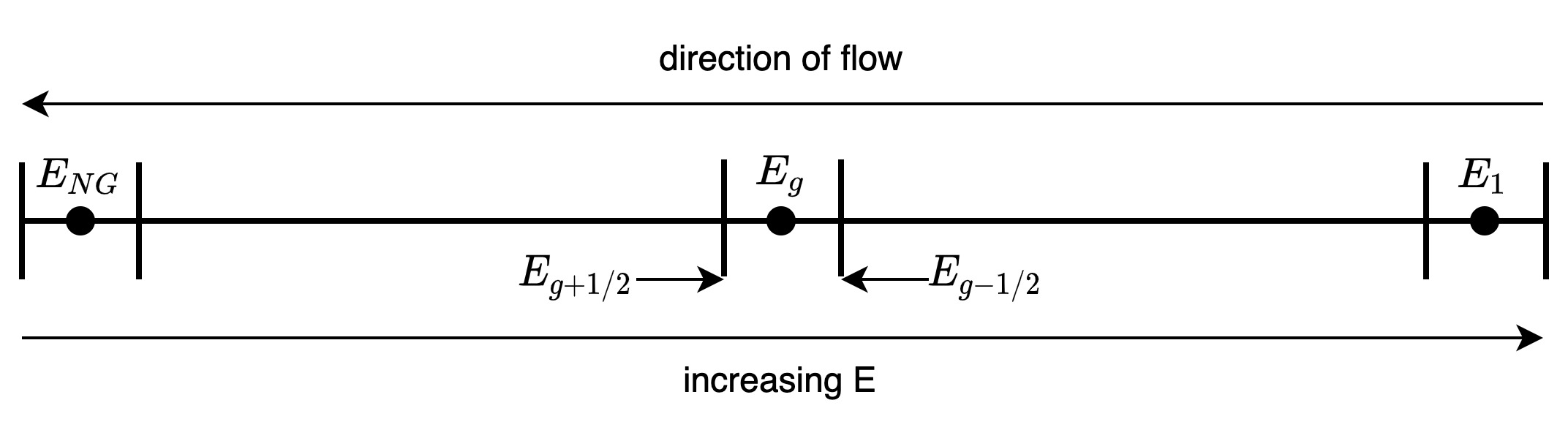}
	\caption{Energy domain discretization}
	\label{fig:energy_discretization}
\end{figure}

The computational domain of the equation is given as
$\mathscr{D} = (0, z_{max}) \cross (E_{min}, E_{max}), \mathscr{D} \subset
\mathbb{R}^2$. The solution of the one-dimensional Fokker-Planck equation
is the Fokker-Planck flux $\varphi_{FP}(z, E) \in \mathscr{H}$
where $\mathscr{H} = \mathscr{L}_2(\mathbb{R}^2) $ is a real Hilbert
space with an associated inner product defined as
\begin{align*}
  \langle f, g \rangle =
	\int\limits_0^{\infty} \dd z \int\limits_0^{\infty} \dd E f g.
\end{align*}
To ensure a unique solution to Equation \ref{eq:1DFP} boundary conditions
must be imposed, namely
\begin{align}
	\text{BCE: }& \eval{\varphi_{FP}(z, E)}_{E=E_{max}} = 0,
	 						\eval{\pdv{\varphi_{FP}(z, E)}{E}}_{E=E_{max}} = 0.
							\label{eq:1DFP_BCE}\\
	\text{BCS: }& \varphi_{FP}(0, E) = A e^{-\qty(\frac{E - E_0}{\sigma_E})^2}.
							\label{eq:1DFP_BCS}
\end{align}
The boundary conditions in energy (BCE) are homogeneous
Dirichlet and Neumann conditions while the boundary condition
in space (BCS) is given by a Gaussian function in energy.
Gerbershagen \cite{gerbershagenSimulationsMeasurementsProton2017} showed that
this is a realistic energy spectrum for a proton beam that has suffered
energy degradation. In line with usual practice, a rigorous proof of the
existence and uniqueness of the solution to Equation \ref{eq:1DFP} and its
associated boundary conditions is not given and these properties are assumed
to be true.
The energy component of the domain $\mathscr{D}$ is discretized into a number
$NG$ of groups. In a given group $g$ the high energy boundary is denoted by
$E_{g-1/2}$, the low energy one by $E_{g+1/2}$ and the center value by
$E_g$. Thus, $E_{max} = E_{1/2}$ and $E_{min} = E_{NG + 1/2}$. An illustration
of this discretization can be seen in Figure \ref{fig:energy_discretization}.

The spatial part of the domain
$\mathscr{D}$ is discretized into a number of steps $N_s$ with the interval
length $\Delta z$ allowed to vary on a per step basis and the start and end
points of the spatial domain are given by $z_0 = \num{0} $ and
$z_{Ns} = z_{max}$.

\subsection{Semi-discrete variational formulation}

The first step to obtain an approximation to the solution of Equation
\ref{eq:1DFP} and its associated boundary conditions
\ref{eq:1DFP_BCE}, \ref{eq:1DFP_BCS} is to obtain the semi-discrete
variational formulation. To do so, several quantities
must be defined. First, the jump and the average of the flux at the
edges of an energy group are defined as
\begin{align*}
	[\varphi] &= \varphi(E^-_{j}) - \varphi(E^+_{j}), \\
	\text{ and }
	\qty{\varphi} &= \frac{1}{2} \qty(\varphi(E^-_{j}) + \varphi(E^+_{j}) ),
\end{align*}
where $j = \frac{1}{2}, \ldots, NG + \frac{1}{2}$ and with $ E^-_{j} =
\lim\limits_{\epsilon \downarrow 0} (E_{j} - \epsilon) $ and $E^+_{j} =
\lim\limits_{\epsilon \downarrow 0} (E_{j} + \epsilon)$. Special cases are
defined at the boundary of the energy domain where
\begin{alignat}{2}
	[v(E_{NG + 1/2})] &= - v(E_{NG + 1/2}^+) \text{, }
	&& \qty{v(E_{NG + 1/2})} = v(E_{NG + 1/2}^+) \text{, and} \nonumber \\
	[v(E_{1/2})] &=  v(E_{1/2}^-) \text{, }
	&&\qty{v(E_{1/2})} = v(E_{1/2}^-) \nonumber.
\end{alignat}
Second, the penalty term is defined as
\begin{align*}
	J_0(v, w) = \mathlarger\sum\limits_{j=1/2}^{NG+1/2}
							\frac{\sigma^0}{h_{j-1,j}} [v(E_j)][w(E_j)]
\end{align*}
where $h_{j-1,j} = \max(\Delta E_{j-1}, \Delta E_{j})$ and $\sigma^0$ is a real
and nonnegative number bounded from below. The role of this term is to penalize
the jumps in the solution.

By multiplying Equation \ref{eq:1DFP} with a test function $v$,
integrating over one group, thereafter summing over all energy groups
and making use of the definitions of the jump and the average, the semi-
discrete variational formulation is found to be
\begin{align}
	\int\limits_{E_{min}}^{E_{max}} \dd E \pdv{\varphi_{FP}}{z} v
	+ a_{SIPG}(\varphi_{FP}, v)
	- \int\limits_{E_{min}}^{E_{max}} \dd E \pdv{S^* \varphi_{FP}}{E} v
	+ \int\limits_{E_{min}}^{E_{max}} \dd E \Sigma_a \varphi_{FP} v
	= 0,
	\label{eq:semi_discr_var_form}
\end{align}
where the SIPG bilinear $a_{SIPG}$ is
\cite{riviereDiscontinuousGalerkinMethods2008}
\begin{equation}
\label{eq:sipg_bilinear}
	a_{SIPG}(\varphi_{FP}, v) = \int\limits_{E_{min}}^{E_{max}} T
	 														 \pdv{\varphi_{FP}}{E} \dv{v}{E} \dd E
														 + \sum\limits_{\Gamma_i}
														 - \qty{T \pdv{\varphi_{FP}}{E}} \cdot [v]
														 - [\varphi_{FP}] \cdot \qty{T \dv{v}{E}}
														 + \frac{\sigma}{\Delta E} [\varphi_{FP}][v],
\end{equation}
where $\Gamma_i$ denotes the interior points of the energy domain. Following
Hillewaert's work \cite{hillewaertDevelopmentDiscontinuousGalerkin2013},
the penalty parameter was set as a function of the maximum polynomial degree
$\max(\deg(p^i_g))$ of the basis functions, namely
\begin{equation}
	\sigma^0 = \frac{(\max(\deg(p^i_g)) + 1)^2}{2}.
\end{equation}

Both a coercivity analysis and the proof of equivalence between the
semi-discrete variational formulation from Equation \ref{eq:semi_discr_var_form}
and the model problem \ref{eq:1DFP} with its
associated boundary conditions \ref{eq:1DFP_BCE} and \ref{eq:1DFP_BCS} are
beyond the scope of this paper and can be found in the work of
Hillewaert and Riviere respectively
\cite{hillewaertDevelopmentDiscontinuousGalerkin2013,
riviereDiscontinuousGalerkinMethods2008}.

\subsection{Basis functions}

The first three group-centered Legendre polynomials
\begin{align}
	\label{eq:energy_basis_funcs}
	p^i_g(E) \equiv P_i\qty(\frac{2}{\Delta E_g} (E - E_g)), i=0,1,2
\end{align}
were used as the basis functions for the expansion of the flux in the
computational domain as
\begin{align}
	\varphi_{FP}(z,E) = \sum\limits_{g=1}^{NG} \sum\limits_{i=0}^2
											\varphi^i_g(z) p^i_g(E).
	\label{eq:phi_FP_expansion}
\end{align}
Introducing the expansion from Equation \ref{eq:phi_FP_expansion} into the
semi-discrete variational formulation from Equation \ref{eq:semi_discr_var_form}
and sequentially replacing the function $v$ with the chosen basis functions
$p^i_g(E)$ yields a system of linear equations. This system can be written as
\begin{align}
	M \dv{\vb{\Phi}}{z} + G \vb{\Phi} = 0,
\end{align}
where $\vb{\Phi}$ is a vector with dimension $ (1 + \max(\deg(p^i_g)))
\cross NG $ and its elements are given by the unknown coefficients
$\varphi_g^i(z)$ from Equation \ref{eq:phi_FP_expansion}, the mass matrix M is
a diagonal matrix that in a given group $g$ has elements
$\int \dd E p_g^i(E) p_g^i(E)$ with $i=0,1,2$ along the diagonal and $G$ is
the system matrix which receives contributions from the stopping power,
straggling coefficient and absorption cross section discretization.

This resulting system is discretized in space using the Crank-Nicholson
method. Depending on the chosen number of groups the size of the resulting
system is on the order of $10^3$. This
relatively small size of the system of equations implies that direct
solution methods are comparable in computational time to iterative ones.
To this end, the banded system solver DGBSV from the LAPACK
library \cite{lapack99} was used.

Initially, the algorithm used first order basis functions.
However, the resulting fluxes for coarse energy and spatial grids
resulted in unphysical negative values. Thereafter, second order basis
functions were implemented.

%% file: src/theory_fermi_eyges.tex
\section{The Fermi-Eyges equation}
\label{sec:FE}
This section describes the analytical solution to the Fermi-Eyges equation
and the steps taken to implement it. This solution is based on refinements
brought to Fermi's original theory on the distribution of charged particles
undergoing multiple elastic scattering in their passing through matter.
Authors such as Eyges, Brahme and Asadzadeh
\cite{eygesMultipleScatteringEnergy1948,
brahmeSimpleRelationsPenetration1975,
gebackAnalyticalSolutionsPencilBeam2012a} have
brought the theory into its form presented here. A full derivation from basic
principles is beyond the scope of this document and can be found in the
previously mentioned publications.

The Fermi-Eyges equation
\begin{align}
	\label{eq:FE}
	\Upsilon\qty[\varphi_{FE}] = \pdv{\varphi_{FE}}{z}
	+ \Omega_x \pdv{\varphi_{FE}}{x} + \Omega_y \pdv{\varphi_{FE}}{y}
	- \Sigma_{tr}(z) \qty(\pdv[2]{\varphi_{FE}}{\Omega_x}
	+ \pdv[2]{\varphi_{FE}}{\Omega_y}) = 0
\end{align}
can be solved by separating the x and y directions, namely
$ \varphi_{FE}(\vb*{r},\Omega_x,\Omega_y) = H(z,x,\Omega_x) \cdot
H(z,y,\Omega_y)$. This results in two separate PDEs for each direction
\begin{align}
	\label{eq:FE_split}
	\pdv{H(z,\xi,\omega)}{z} + \omega \pdv{H(z,\xi,\omega)}{\xi}
	- \Sigma_{tr}(z) \pdv[2]{H(z,\xi,\omega)}{\omega} = 0,
\end{align}
where $ \xi $ stands for one of $ x, y $ and $ \omega $ stands for one of
$ \Omega_x, \Omega_y $. The resulting PDEs have the same boundary
condition imposed, namely
\begin{align}
	\label{eq:FE_BC}
	H(0,\xi,\omega) = C \exp\qty(-\qty(a_1\xi^2+a_2\xi\omega+a_3\omega^2)),
\end{align}
with $a_i \in \mathbb{R} ,\forall i=1,2,3$ and $C>0$.
The solution of Equation \ref{eq:FE_split} is found by applying two-dimensional
Fourier transforms in $ \xi $ and $ \omega $ and accounting for the Gaussian
initial condition as detailed by Eyges and Brahme
\cite{brahmeSimpleRelationsPenetration1975,
eygesMultipleScatteringEnergy1948}.
In doing so the solution to the Fermi-Eyges Equation
\ref{eq:FE} is found to be
\begin{align}
	\label{eq:FE_sol}
	\varphi_{FE}(z, \vb*{\rho}, \vu*{\Omega}) = \frac{A^2}{4 \pi^2}
	\frac{\exp\qty(-\frac{|\vb*{\rho}|^2}{2 \overline{\xi^2}(z)})}
	{\overline{\xi^2}(z)} \frac{\exp\qty(-\frac{1}{2 B(z)} \qty|
	\vu*{\Omega} - \frac{\overline{\theta\xi}(z)}
	{\overline{\xi^2}(z)}\vb*{\rho}|^2 )}{B(z)},
\end{align}
where $ \vb*{\rho} = (x, y) \text{, } \vu*{\Omega} = (\Omega_x, \Omega_y) $ and the remaining quantities are defined as
\begin{align*}
	B(z) &= \overline{\theta^2}(z) - \frac{\qty(\overline{\theta\xi}(z))^2}{\overline{\xi^2}(z)} \text{, } A = \frac{2 \pi C}{D}, D = 4 a_1 a_3 - a_2^2.
\end{align*}
Jette \cite{jetteElectronDoseCalculation1988} showed that if there is any
scattering at all, then $B \geq 0$ must hold. This was used as a check that
the obtained coefficient values were not spurious. The coefficients
$\overline{\theta^2} , \overline{\theta\xi}, \overline{\xi^2}$ present in the
Fermi-Eyges solution are the moments of the $ \Sigma_{tr} $ transport cross
section and are found from the following equations
\begin{subequations}
	\label{eq:FE_solved_ODEs}
	\begin{align}
		\overline{\theta^2}(z) &= \overline{\theta^2}(0)
		+ \int\limits_{0}^z \Sigma_{tr}(z') \dd z' \text{, with }
		\overline{\theta^2}(0) = \frac{2 a_3}{D}  \\
		\overline{\theta \xi}(z) &= \overline{\theta \xi}(0)
		+ \overline{\theta^2}(0) z
		+ \int\limits_{0}^z (z-z') \Sigma_{tr}(z') \dd z' \text{, with }
		\overline{\theta\xi}(0) = \frac{a_2}{D} \\
		\overline{\xi^2}(z) &= \overline{\xi^2}(0)
		+ 2 \overline{\theta \xi}(0) z
		+ \overline{\theta^2}(0) z^2
		+  \int\limits_{0}^z (z-z')^2 \Sigma_{tr}(z') \dd z' \text{, with }
		\overline{\xi^2}(0) = \frac{2 a_1}{D}.
	\end{align}
\end{subequations}
where
\begin{align}
	\label{eq:Sigma_tr}
	\Sigma_{tr}(z) = \int\limits_{-1}^{1} \dd \mu
	\Sigma_s(E_a(z), \mu) \qty(1-\mu)  \text{, with }
	\mu=\cos(\vu*{\Omega} \cdot \vu*{\Omega}'),
\end{align}
and $ \Sigma_s $ is the macroscopic elastic scatter cross section. Gottschalk
\cite{gottschalkTechniquesProtonRadiotherapy2012} showed that the FE
coefficients $\overline{\theta^2}(z), \overline{\xi^2}(z),
\overline{\theta \xi}(z)$ can be intepreted as the variances of the angular
direction, the lateral position and the covariance of the lateral position and
the angular direction respectively.

Next to its analytical nature an important feature of the Fermi-Eyges solution
from Equation \ref{eq:FE_sol} is that it is a Gaussian function
in both the spatial and angular directions with coefficients that are
determined by the average depth-dependent beam energy and the elastic scatter
cross section from Equation \ref{eq:Sigma_tr} corresponding to that energy.
A disadvantage of this solution is that only the
average depth dependent energy instead of the full beam energy spectrum is used
to calculate the coefficients. As lower energy protons tend to scatter more
it is expected that only using the average beam energy will result in an
underestimation of the amount of scatter that the proton beam suffers.

\subsection{Solution method}

The coefficients of the boundary condition are chosen in such a way that
Equation \ref{eq:FE_BC} represents the two-dimensional normal distribution.
By setting the average values in $ \xi $ and $ \omega $ to zero
the coefficients $ a_i , i \in [1, 2, 3]$ from Equation \ref{eq:FE_BC} are
easily identified to be equal to
\begin{align}
\label{eq:FE_init_coeffs}
	a_1 = \frac{1}{2(1-\varrho^2) \sigma_\xi^2} \text{, }
	a_2 = -\frac{\varrho}{(1-\varrho^2) \sigma_\xi \sigma_\omega} \text{, }
	a_3 = \frac{1}{2(1-\varrho^2)\sigma_\omega^2},
\end{align}
where $ \varrho $ is the correlation coefficient between the
spatial dimension $ \xi $ and the angular dimension $ \omega $, $ \sigma_\xi $
standard deviation in $ \xi $ and $ \sigma_\omega $ standard deviation in
$ \omega $. The $a_i$ coefficients are thereafter used to initialize the values
of the FE coefficients.

To compute the FE coefficients at a given depth the average beam energy at that
depth must be known. This quantity was defined as
\begin{align}
	E_a(z) = \frac{\int\limits_{0}^\infty \dd E \varphi_{FP}(z, E) E}
								{\int\limits_{0}^\infty \dd E \varphi_{FP}(z, E)}.
\end{align}
The average energy is thereafter introduced into the elastic scatter cross
section via the classical relationship between speed
and energy $v_p = \sqrt{2E_a(z)/m_p}$. The elastic scatter cross section is in
turn used to compute the transport cross section from Equation
\ref{eq:Sigma_tr}. To compute the angle integral the QAGE routine from the
QUADPACK library was used \cite{piessensQUADPACKSubroutinePackage1983}.

Once $ \Sigma_{tr}(z) $ is known for all the points of z dimension, the
coefficients given in Equation \ref{eq:FE_solved_ODEs} can be calculated.
As $z$ increases in the integrals from Equations \ref{eq:FE_solved_ODEs} so
do the integrands and the computational expense of these integrals.
We chose to approximate $\Sigma_{tr}$ in a given step as the average of its
values at the start and endpoint of the step. In doing so, the integrals could
be re-written to depend only on the previous value.

\subsection{The planar integral approximation}
In the response computation the angle integrated FE flux is needed.
In the derivation of the Fermi-Eyges solution the domain of
$ \Omega_x $ and $ \Omega_y $ is extended from their
normal range to the $ (-\infty, \infty) $ range in order to apply the Fourier
transforms. Since the Fermi-Eyges solution is obtained
through this extension, the same extension should be consistently applied
throughout the calculations that involve this solution. Thus, the angular
integral can be approximated to
\begin{align}
	\int\limits_{4 \pi} \exp\qty(-\frac{1}{2 B(z)}
	\qty| \vb*{\Theta} - \frac{\overline{\theta\xi}(z)}
	{\overline{\xi^2}(z)}\vb*{\rho}|^2 )
	\sin \theta \dd \theta \dd \phi
	& \approx \int\limits_{-\infty}^{\infty} \int\limits_{-\infty}^{\infty}
	\exp\qty(-\frac{1}{2 B(z)} \qty| \vb*{\Theta}
	- \frac{\overline{\theta\xi}(z)}{\overline{\xi^2}(z)}\vb*{\rho}|^2 )
	\dd \Omega_x \dd \Omega_y \nonumber \\
	& = \int\limits_{-\infty}^{\infty} \int\limits_{-\infty}^{\infty}
	\exp\qty(-\frac{1}{2 B(z)} \qty[\qty(\Omega_x - c_x)^2
	+ \qty(\Omega_y - c_y)^2]) \dd \Omega_x \dd \Omega_y \nonumber \\
	& = \sqrt{2 \pi B(z)} \cdot \sqrt{2 \pi B(z)} = 2 \pi B(z).
\end{align}
Thus, the angularly integrated FE flux is
\begin{align}
	\label{eq:int_4pi_phi_FE}
	\Psi_{FE}(x,y,z) = \int\limits_{4 \pi} \varphi_{FE}(x,y,z,\Omega_x,\Omega_y)
	 \dd \vu*{\Omega}
	 = \frac{A^2}{2 \pi} \frac{1}{\overline{\xi^2}(z)}
	 \exp\qty(-\frac{|\vb*{\rho}|^2}{2 \overline{\xi^2}(z)}).
\end{align}

\subsection{Data sources}

\begin{figure}[!h]
  \vspace{-.4cm}
  \centering
  \begin{minipage}{0.49\textwidth}
		\centering
		\includegraphics[width=\linewidth]{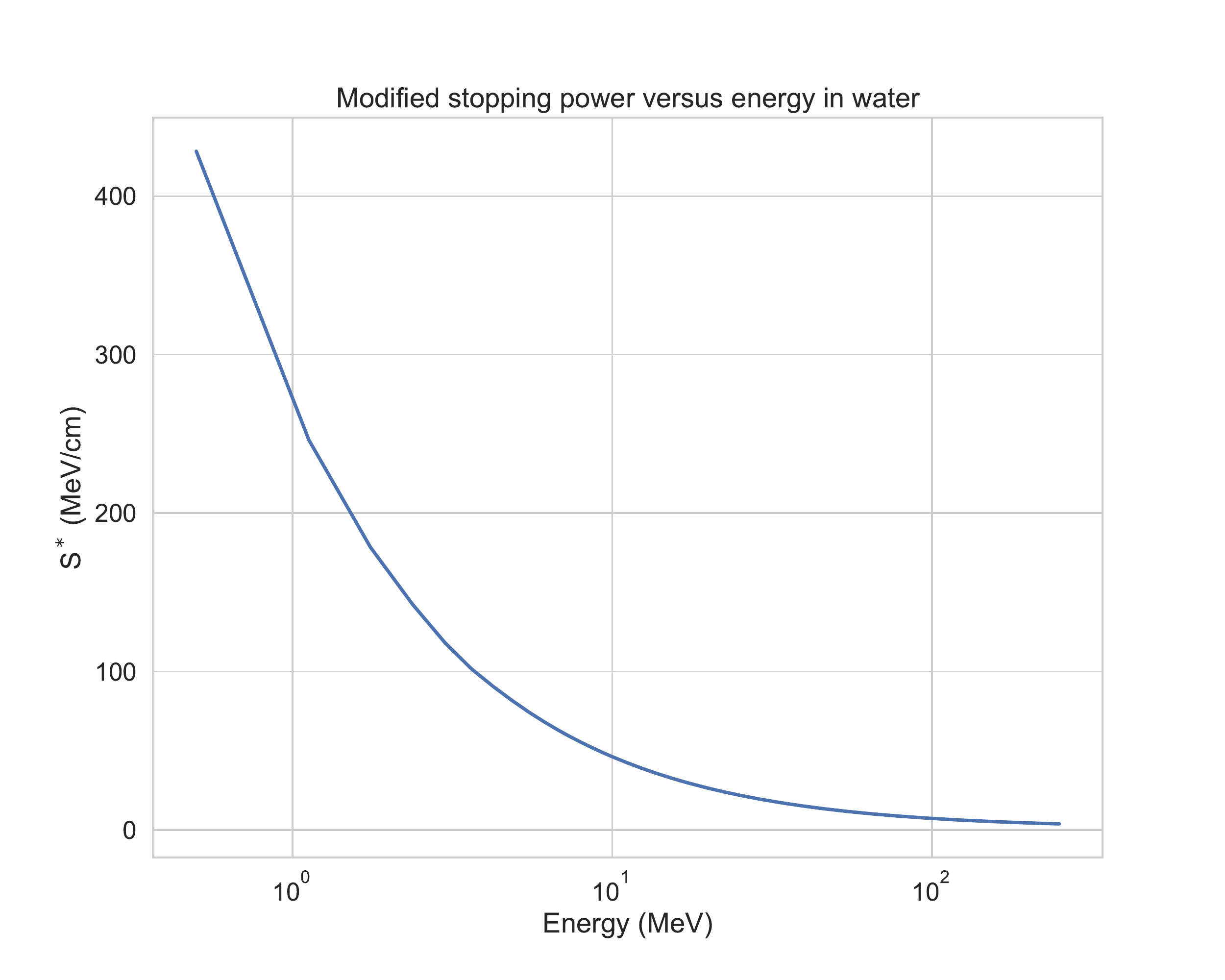}
		\caption{Water stopping power versus energy}
    \label{fig:S_vs_E_in_water}
  \end{minipage}\hfill
  \begin{minipage}{0.49\textwidth}
    \centering
		\includegraphics[width=\linewidth]{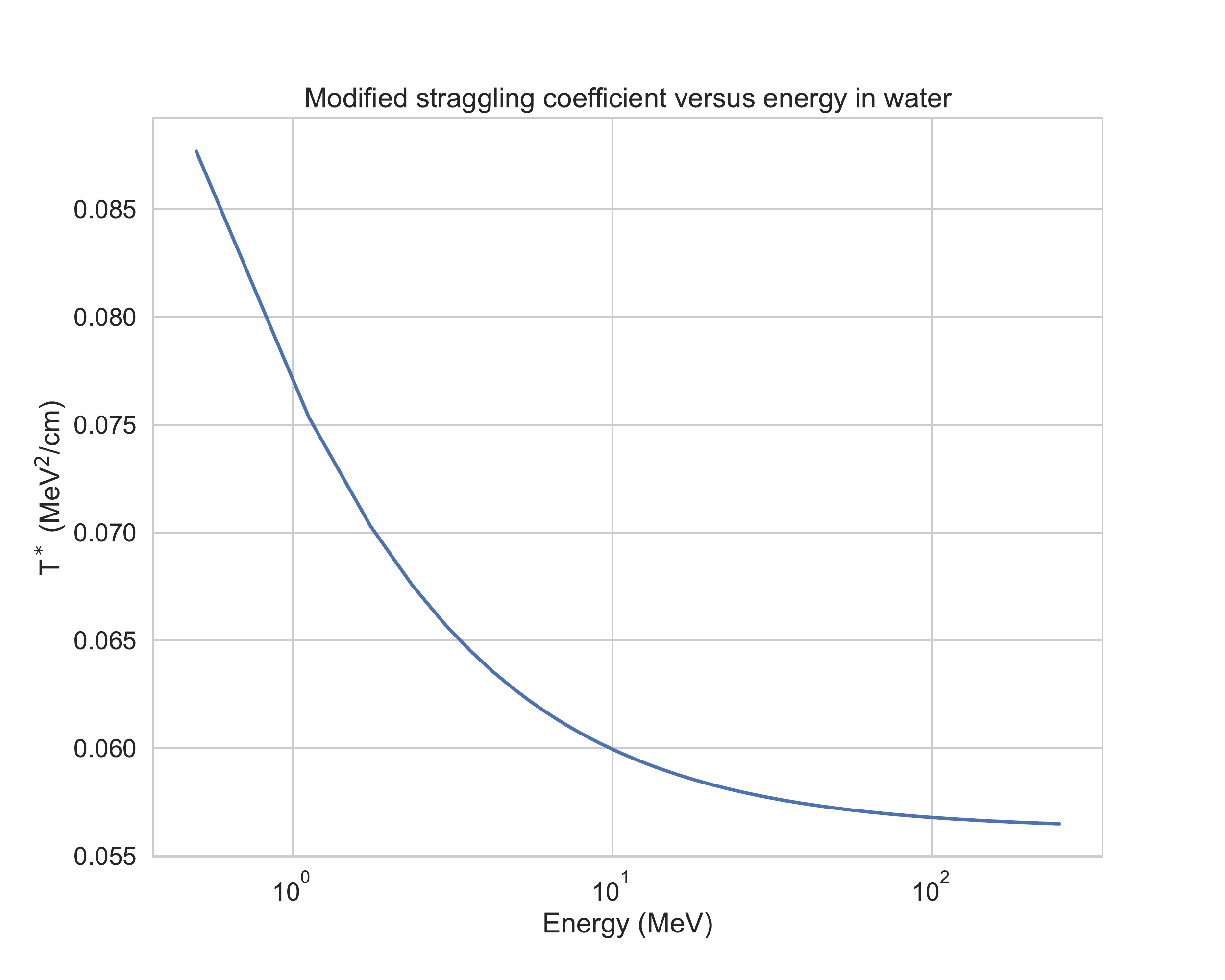}
		\caption{Water straggling coefficient versus energy}
    \label{fig:T_vs_E_in_water}
  \end{minipage}
\end{figure}

In order to obtain the solution to the two PDEs and the response,
the stopping power, straggling coefficient, absorption cross section
and elastic scatter cross section must be known
as a function of energy and tissue composition. The CT scan HU values were
converted to density and fractional compositions according to Schneider's method
\cite{schneiderCorrelationCTNumbers2000}. The density and fractional
composition were used to interpolate nuclide specific tables of the
stopping power versus energy. The tables were extracted from TOPAS
\cite{perlTOPASInnovativeProton2012a} using an adapted extension
distributed on the TOPAS forum \cite{schuemannStoppingPowerExtraction}. The
stopping power for protons in water versus energy can be seen in Figure
\ref{fig:S_vs_E_in_water}.

The straggling coefficient represents the statistical variation around the
mean of the energy loss of a proton in a material. The consequence of energy
straggling is the spreading of the energy spectrum of an initially
mono-energetic beam \cite{noshadInvestigationEnergyStraggling2012}.
The equation that was used for the straggling coefficient is
\cite{williamsPassageParticlesMatter1932}
\begin{align}
	T(E, N_\mathbb{A}) =
	\mathlarger\sum\limits_{i \in \mathbb{A}} \frac{1}{(4 \pi \epsilon_0)^2}
						  N_i 4 \pi e^4 Z_i \qty(1 + \frac{4 I_i}{3 m_e v_p^2}
							\ln \frac{2 m_e v_p^2}{I_i}),
\end{align}
where $N_\mathbb{A}$ is the set of atomic densities corresponding to the
set of atoms atoms $\mathbb{A}$ that were
considered to constitute human tissue, namely
$\mathbb{A} = \qty{\text{H, C, N, O, Na, Mg, P, S, Cl, Ar, K, Ca}}$.
Moreover, $Z_i$ is the atomic number of the
target atom $i$ with $ i \in \mathbb{A}$, $\epsilon_0$ is the vacuum permitivity
constant, $e$ is the elementary charge, $m_e$ is the electron mass,
$v_p$ is the proton speed, $I_i$ is the mean atomic excitation energy of atom
$i$. The straggling coefficient for protons versus energy in water can be
seen in Figure \ref{fig:T_vs_E_in_water}.


The elastic scatter cross section can be found by considering the deflection
that a proton suffers due to the Coulomb field of the nucleus. A derivation of
this can be found in the work of Goldstein
\cite{goldsteinClassicalMechanics2002} who gives the microscopic elastic
scatter cross section for protons incident on a target nucles $t, t \in
\mathbb{A}$ with atomic number $Z_t$ and atomic mass numbers $ A_t $ as
\begin{align}
	\label{eq:Sigma_s}
	\sigma_{s,t}(E, \mu) = \frac{\qty( 1 + \frac{2 \mu}{A_t}
										 + \frac{1}{A_t^2})^{3/2}}{1+\frac{\mu}{A_t}}
	\qty(\frac{Z_t e^2}{4 \pi \epsilon_0 m_0 v_p^2})^2
	\frac{1}{\qty(1 - \mu + 2\eta)^2},
\end{align}
where $ m_0 $ is the reduced mass which is defined by
\begin{equation*}
	\frac{1}{m_0} = \frac{1}{m_p} + \frac{1}{m_t}
\end{equation*}
with $ m_p $ the mass of the proton and $ m_t $ the mass of the target nucleus,
$ v_p $ is the incident speed of the proton, $ \epsilon_0 $ is the vacuum
permittivity, $ e $ is the elementary charge and
\begin{equation*}
	\eta = \Theta_{min}^2 = \qty( \frac{Z_t^{1/3} \alpha m_e c}{p} )^2
\end{equation*}
with $ m_e $ the electron mass, $ \alpha $ the fine structure constant, $ c $
the speed of light and $ p $ the momentum of the incident proton. Equation
\ref{eq:Sigma_s} is used to define the macroscopic scatter cross section as
\begin{align}
	\Sigma_s(E, \mu, N_\mathbb{A}) = \sum\limits_{i \in \mathbb{A}}
																    N_i \sigma_{s,i}(E, \mu)
\end{align}
with $N_i, i \in \mathbb{A}$ the individual atomic density in the
material under consideration.

\begin{figure}[!h]
	\vspace{-.4cm}
	\centering
	\includegraphics[width=.6\linewidth]{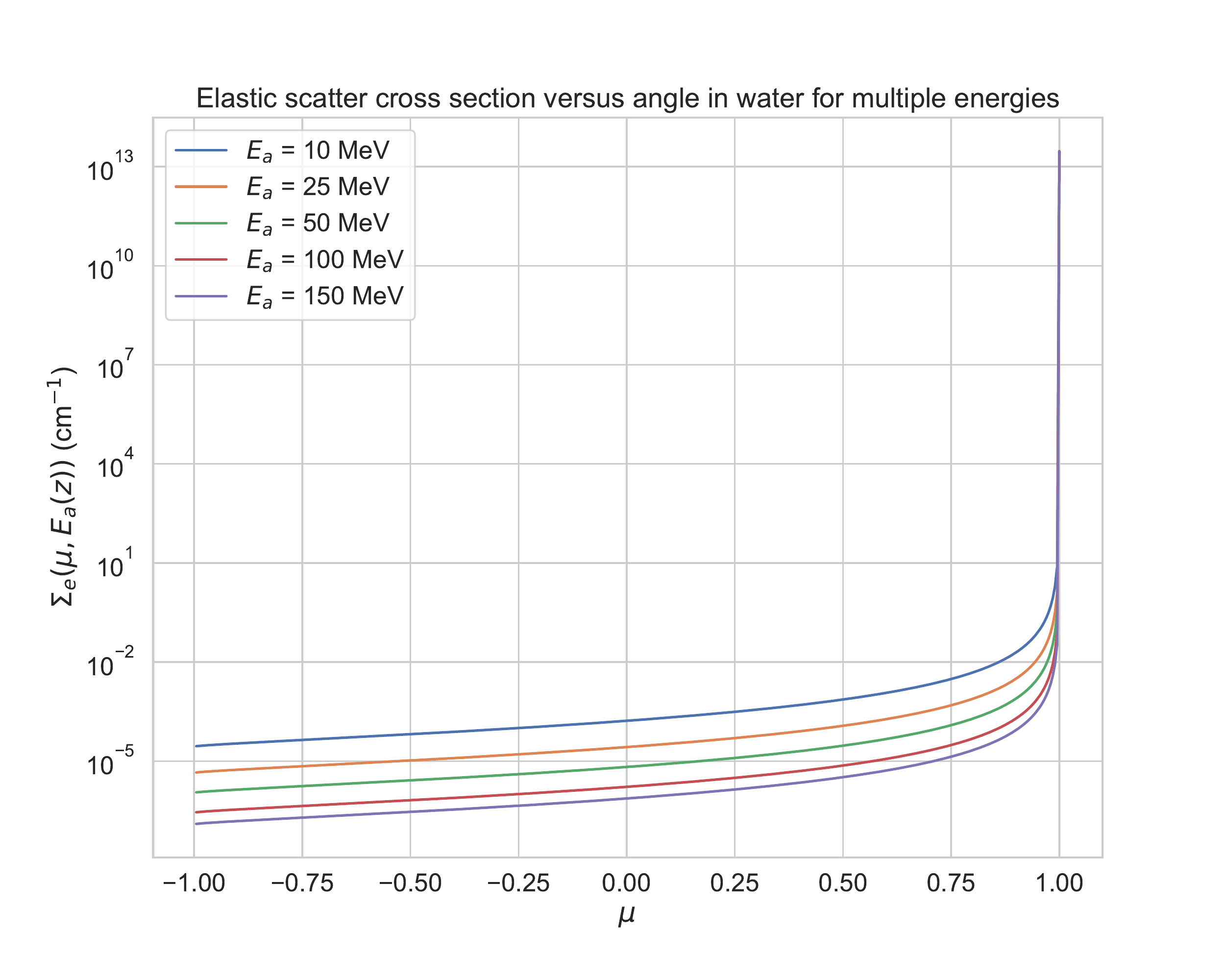}
	\caption{Water elastic scatter cross section for protons versus angle for multiple energies}
\end{figure}


%% file: src/theory_adjoint.tex
\section{Response change}
\label{sec:response}
Using the Fokker-Planck flux $\varphi_{FP}$, the FE coefficients from Equation
\ref{eq:FE_solved_ODEs} and a given set of system parameters $\vb*{\alpha}$
the deposited energy in an arbitrary ROI is computed via
\begin{align*}
	R(\vb*{\alpha}, \varphi_{FP})
	&= - \int\limits_{ROI} \dd V \int\limits_{4 \pi} \dd \vu*{\Omega}
   		 \int\limits_{E_{min}}^{E_{max}} \dd E \qty [E \pdv{S \varphi}{E}
     + E \pdv{E} \qty(T \pdv{\varphi}{E}) - E \Sigma_a \varphi].
 \end{align*}
It is of interest to assess how the response changes depending on changes in the
system parameters $\vb*{\alpha}$. This change in the response can be
described as a direct and an indirect change. The direct change is the one that
results from the change in the system parameters being directly used to compute
the response. The indirect change comes through the FP flux and FE coefficients
which are perturbed when changes in the system parameters are present.
Thus, for each new vector of system parameters $\vb*{\alpha}$ a new solution to
the Fokker-Planck and Fermi-Eyges systems must be obtained.

This section describes the functional relationship between the change in the
response $\var{R}$ and the changes in the system parameters
$\var{\vb*{\alpha}}$ and the FP flux $\var{\varphi_{FP}}$. Moreover, it
describes a methodology that allows cheaply evaluating the desired change
in the response without re-computing the Fokker-Planck flux and the Fermi-Eyges
coefficients.

\subsection{The change in the response}

The response can be written using the ansatz from Equation
\ref{eq:flux_ansatz} as
\begin{align}
\label{eq:R_def}
	  R(\vb*{\alpha}, \varphi_{FP})& = - \int\limits_{ROI} \dd V
          \int\limits_{4 \pi} \dd \vu*{\Omega} \varphi_{FE}
	         \int\limits_{E_{min}}^{E_{max}} \dd E \qty[
          E \pdv{S \varphi_{FP}}{E}
        + E \pdv{E} \qty(T \pdv{\varphi_{FP}}{E})
        - E \Sigma_a \varphi_{FP}].
\end{align}
Making use of the definition from Equation \ref{eq:int_4pi_phi_FE}
for the $4 \pi$ integrated Fermi-Eyges flux allows writing the response
from Equation \ref{eq:R_def} in a shortened form
\begin{align}
	R(\vb*{\alpha}, \varphi_{FP}) =
	\int\limits_{ROI} \dd V \Psi_{FE}(\vb{r})
													D_{FP}(z, \varphi_{FP}, \vb*{\alpha}),
	\label{eq:short_form_R}
\end{align}
where
\begin{align*}
D_{FP}(z, \varphi_{FP}, \vb*{\alpha}) &=
\qty[E S \varphi_{FP}\eval{}_{E_{min}}
				+ \int\limits_{E_{min}}^{E_{max}} \dd E S \varphi_{FP}
				+ \int\limits_{E_{min}}^{E_{max}} \dd E T \pdv{\varphi_{FP}}{E}
				+ \sum\limits_{\Gamma_i} - \qty[\varphi] T
				+ \int\limits_{E_{min}}^{E_{max}} \dd E E \Sigma_a \varphi_{FP}].
\end{align*}
The change in the response due to changes in the system parameters can be
found by computing the Gateaux-differential. Let $\vb*{e^0}
= (\vb*{\alpha}, \varphi_{FP})$ and $\vb{h} = (\var{\vb*{\alpha}},
\var{\varphi_{FP}})$. Then,
\begin{align}
	\var{R}(\vb*{e^0}, \vb*{h})
	&= \dv{t} \eval{R(\vb*{e^0} + t \vb*{h})}_{t=0}
	= \int\limits_{ROI} \dd V \qty[\var{\Psi_{FE}(\vb{r})} D_{FP}(z)
	+  \Psi_{FE}(\vb{r}) \var{D_{FP}(z)} ].
	\label{eq:generalied_var_R}
\end{align}
The Gateaux-differential of $D_{FP}$ is computed to be
\begin{align*}
\var{D_{FP}} &= \qty{E \var{S} \varphi_{FP}\eval{}_{E_{min}}
			+ \int\limits_{E_{min}}^{E_{max}} \dd E \var{S} \varphi_{FP}
			+ \int\limits_{E_{min}}^{E_{max}} \dd E \var{T} \pdv{\varphi_{FP}}{E}
			+ \sum\limits_{\Gamma_i} - \qty[\varphi] \var{T}
			+ \int\limits_{E_{min}}^{E_{max}} \dd E E \var{\Sigma_a} \varphi_{FP}} \\
			&+ \qty{E S \var{\varphi_{FP}}\eval{}_{E_{min}}
			+ \int\limits_{E_{min}}^{E_{max}} \dd E S \var{\varphi_{FP}}
			+ \int\limits_{E_{min}}^{E_{max}} \dd E T \pdv{\var{\varphi_{FP}}}{E}
			+ \sum\limits_{\Gamma_i} - \qty[\var{\varphi}] T
			+ \int\limits_{E_{min}}^{E_{max}} \dd E E \Sigma_a \var{\varphi_{FP}} } \\
			&= \var{D_{FP,dir}}(\var{\vb*{\alpha}}, \varphi_{FP})
			+ \var{D_{FP,indir}}(\vb*{\alpha}, \var{\varphi_{FP}})
\end{align*}
where $\var{D_{FP,dir}}(\var{\vb*{\alpha}}, \varphi_{FP})$
is the direct change in the response due to the change in the system parameter
vector and $\var{D_{FP,indir}}(\vb*{\alpha}, \var{\varphi_{FP}})$
is the indirect change in the
response due to the perturbation in the dependent system variable.

The Gateaux-differential of $\Psi_{FE}(\vb*{r})$ is found to be
\begin{align}
	\var{\Psi_{FE}}(\vb*{r}) &= \dv{t} \Psi_{FE}(A+t\var{A},
	\overline{\xi^2}(z) + t \var{\overline{\xi^2}(z)})\eval{}_{t=0}
	= \Psi_{FE}
	\qty[\frac{2 \var{A}}{A} - \frac{\var{\overline{\xi^2}(z)}}
	{\overline{\xi^2}(z)} + \frac{x^2 + y^2}{2}
	\frac{\var{\overline{\xi^2}(z)}}{\overline{\xi^2}(z)^2} ].
\end{align}
The quantity $ \var{\Psi_{FE}}(\vb*{r}) $ is thereafter laterally integrated
over the X and Y extents of the ROI, namely
\begin{align}
	\iint\limits_{ROI_{XY}} \dd x \dd y \var{\Psi_{FE}} &=
	\iint\limits_{ROI_{XY}} \dd x \dd y \Psi_{FE}
	\qty[\frac{2 \var{A}}{A} - \frac{\var{\overline{\xi^2}(z)}}
	{\overline{\xi^2}(z)} + \frac{x^2 + y^2}{2}
	\frac{\var{\overline{\xi^2}(z)}}{\overline{\xi^2}(z)^2} ] \nonumber \\
	&= \qty[\frac{2 \var{A}}{A} - \frac{\var{\overline{\xi^2}(z)}}
	{\overline{\xi^2}(z)}]
	\underbrace{\iint\limits_{ROI_{XY}} \dd x \dd y \Psi_{FE} }_{I_{f1}}
	+ \frac{\var{\overline{\xi^2}(z)}}{\overline{\xi^2}(z)^2}
	\underbrace{\iint\limits_{ROI_{XY}} \dd x \dd y
	\frac{x^2 + y^2}{2} \Psi_{FE}}_{I_{f2}} \nonumber \\
	&= \qty[\frac{2 \var{A}}{A}
	  - \frac{\var{\overline{\xi^2}(z)}}{\overline{\xi^2}(z)}] I_{f,1}(z)
		+ \frac{\var{\overline{\xi^2}(z)}}{\overline{\xi^2}(z)^2} I_{f,2}(z).
\end{align}
At this point the term $\var{A}$ is set to zero. This term is only non-zero
when the Fermi-Eyges initial condition is perturbed. For the
purpose of this work, no such perturbation was included.
Thus, the laterally integrated $\var{\Psi_{FE}}$ is given as
\begin{align}
	\iint\limits_{ROI_{XY}} \dd x \dd y \var{\Psi_{FE}}
	&= \var{\overline{\xi^2}(z)} \qty[- \frac{I_{f,1}(z)}{\overline{\xi^2}(z)} + \frac{I_{f,2}(z)}{\overline{\xi^2}(z)^2} ].
	\label{eq:int_ROI_var_Psi_FE}
\end{align}

The term $ \var{\overline{\xi^2}(z)} $ contains the unknown
$ \var{\varphi_{FP}} $. Continuing the Gateaux-differential process results in
\begin{align}
	\var{\overline{\xi^2}}(z) &=
	\int\limits_{0}^z (z-z')^2 \var{\Sigma_{tr}(z')} \dd z'
	\text{ where }
  \var{\Sigma_{tr}(z)} =
	\int\limits_{-1}^{1} \var{\Sigma_s(E_a(z), \mu)} \qty(1-\mu) \dd \mu.
	\label{eq:var_xi_def}
\end{align}
As the system parameters change, so does the Fokker-Planck flux
$\varphi_{FP}$. This in turn results in a change in the average depth-dependent
energy $E_a(z)$ of the beam which in turn results ultimately in changes in the
FE coefficients. To compute the effect of a change in the FP flux on the FE
coefficients, the elastic scatter cross section can be re-written to illustrate
the energy dependence by using the classical kinetic energy relationship between
speed and energy as
\begin{align*}
	\Sigma_s(E_a(z), \mu, N_\mathbb{A}) &=
	\mathlarger{\sum}\limits_{i \in \mathbb{A}} N_i
	F_1(\mu, A_i) F_2(Z_i, m_{0i}) \frac{1}{E_a^2}
	\frac{1}{\qty(1 - \mu + \frac{2 c_{\eta, i}}{E_a})^2}, \nonumber \\
	\text{ where }
	F_1(\mu, A_i) &=
	\frac{\qty(1 + \frac{2 \mu}{A_i} + \frac{1}{A_i^2})^{3/2}}{1+\frac{\mu}{A_i}}
	\text{ , }
	F_2(Z_i, m_{0i}) = \qty(\frac{Z_i e^2 m_p}{8 \pi \epsilon_0 m_{0i}})^2
	\text{ and }
	c_{\eta,i} = E_a \eta_i.
\end{align*}
The Gateaux-differential of the elastic scatter cross section is
\begin{align}
	\var{\Sigma_s(E_a(z), \mu, N_\mathbb{A})} &=
	\dv{t} \mathlarger{\sum}\limits_{i \in \mathbb{A}}
	(N_i + t \var{N_i})  F_1(\mu, A_i) F_2(Z_i, m_{0i})
	\frac{1}{(E_a(z) + t \var{E_a(z)})^2}
	\frac{1}{\qty(1 - \mu + \frac{2 c_{\eta,i}}{E_a(z) + t \var{E_a(z)}})^2}
	 \eval{}_{t=0} \nonumber \\
	&=
	\mathlarger{\sum}\limits_{i \in \mathbb{A}}
	\var{N_i} F_1(\mu, A_i) F_2(Z_i, m_{0i})
	\frac{1}{E_a(z)^2}
	\frac{1}{\qty(1 - \mu + \frac{2 c_{\eta,i}}{E_a(z)})^2} \nonumber \\
	&+ \mathlarger{\sum}\limits_{i \in \mathbb{A}}
	N_i F_1(\mu, A_i) F_2(Z_i, m_{0i})
	\frac{\var{E_a(z)}}{E_a(z)^2}
	\frac{1}{\qty(1 - \mu + \frac{2 c_{\eta,i}}{E_a(z)})^2}
	\qty[ \frac{-2}{E_a(z)}
	    + \frac{4 c_{\eta, i} }{E_a^2(z)}
			  \frac{1}{1 - \mu + \frac{2 c_{\eta, i}}{E_a(z)}}] \nonumber \\
	&= \var{\Sigma_{s1}}(\var{N_\mathbb{A}}) + \var{\Sigma_{s2}}(\var{E_a(z)}),
	\label{eq:var_Sigma_s}
\end{align}
where $\var{N_\mathbb{A}}$ is the set of perturbations in the atomic density
set previously defined as $N_\mathbb{A}$. It can be seen that here as well the
change in the elastic scatter cross
section can be described as a direct change due to the atomic composition
in the tissue $\var{N_\mathbb{A}}$ and the change due to the change in the beam
energy spectrum. The last component is to obtain the relationship between the
change in the energy spectrum $\var{E_a(z)}$ and the change in the flux itself
$\var{\varphi_{FP}}$. The Gateaux-differential of the average
depth-dependent energy is
\begin{align}
	\label{eq:var_E_a}
	\var{E_a(z)}
	= \frac{1}{N_p(z)} \int_0^\infty \dd E E \var{\varphi_{FP}
	- \frac{E_b(z)}{N_p(z)^2} \int_0^\infty \dd E \var{\varphi_{FP}}},
\end{align}
where the number of particles and the beam energy at a given depth are defined
as
\begin{align}
	N_p(z) &= \int\limits_0^\infty \dd E \varphi_{FP}(z, E) \text{ and } \\
	E_b(z) &= \int\limits_0^\infty \dd E E \varphi_{FP}(z, E).
\end{align}
At this point the functional relationship between $\var{R}$ and the changes in
$\var{\vb*{\alpha}}$ and $\var{\varphi_{FP}}$ can be obtained. The first
step is to introduce the result from Equation \ref{eq:var_Sigma_s} into
the Gateaux-differential of $\overline{\xi^2}$ given in Equation
\ref{eq:var_xi_def}, yielding
\begin{align}
	\var{\overline{\xi^2}(z)}
	&= \int\limits_0^z \dd z' (z-z')^2\var{\sigma_{tr}(z')}
	 = \int\limits_0^z \dd z' (z-z')^2
		 \int\limits_{-1}^{1} \dd \mu
		 \qty[\var{\Sigma_{s1}}(\var{N_\mathbb{A}}) +
		 \var{\Sigma_{s2}}(\var{E_a(z)}) ] (1-\mu) \nonumber \\
	&= \var{\overline{\xi^2}}_1(\var{N_\mathbb{A}})
	 + \var{\overline{\xi^2}}_2(\var{E_a(z)}).
  \label{eq:var_xi}
\end{align}
The term $ \var{\overline{\xi^2}}_1 $ does not contain any dependencies on the
unknown $ \var{\varphi_{FP}} $ and contributes to the direct effect.
The term $ \var{\overline{\xi^2}}_2 $ does on the other hand contain a
dependency on $ \var{\varphi_{FP}} $. Its $ \var{\varphi_{FP}} $ dependency
is obtained by using Equation \ref{eq:var_E_a}, namely
\begin{align}
	\var{\overline{\xi^2}}_2 &=
	\int\limits_{0}^z  \dd z' (z-z')^2 \nonumber \\
	& \hspace{.3 cm} \underbrace{\int\limits_{-1}^{1} \dd \mu (1-\mu)
	\sum\limits_{i \in \mathbb{A}} N_i F_1(\mu, A_i) F_2(Z_i, m_{0i})
	\frac{\var{E_a(z')}}{E_a(z')^2}
	\frac{1}{\qty(1 - \mu + \frac{2 c_{\eta, i}}{E_a(z')})^2}
  \qty[ \frac{-2}{E_a(z')} + \frac{4 c_{\eta, i} }{E_a^2(z')}
	\frac{1}{1 - \mu + \frac{2 c_{\eta i}}{E_a(z')}}]}_{\var{E_a}(z')
	\cdot I_\mu(z')} \nonumber \\
	&= \int\limits_{0}^z  \dd z' (z-z')^2 \var{E_a}(z') I_\mu(z')
	 = \int\limits_{0}^z  \dd z' (z-z')^2 I_\mu(z')
	    \int_0^\infty \dd E
			\qty(\frac{E}{N_p(z')} - \frac{E_b(z')}{N_p(z')^2})
		  \var{\varphi_{FP}(z', E)}. \nonumber
\end{align}

The next step is to introduce the expression from Equation \ref{eq:var_xi}
into the lateral ROI integration of $ \var{\Psi_{FE	}} $ from Equation
\ref{eq:int_ROI_var_Psi_FE}. For simplicity of notation, let
\begin{align*}
	\Psi^{ROI_{xy}}_{FE}(z) = \int\limits_{ROI_{xy}(z)} \dd x \dd y \Psi_{FE}.
\end{align*}
Using this, $\var{R}$ becomes
\begin{align}
\var{R} &= \int\limits_{ROI_z} \dd z  D_{FP}(z)
	 \int\limits_{ROI_{xy}} \dd x \dd y \var{\Psi_{FE}}
 + \int\limits_{ROI_z} \dd z \Psi^{ROI_{xy}}_{FE}(z) \var{D_{FP}(z)}.
\end{align}
Introducing in this expression the direct and indirect contributions results
in
\begin{align}
\var{R} &= \int\limits_{ROI_z} \dd z \qty {D_{FP}(z)
 													 \var{\overline{\xi^2}(z)}
													 \qty[- \frac{I_{f,1}(z)}{\overline{\xi^2}(z)}
													 			+ \frac{I_{f,2}(z)}{\overline{\xi^2}(z)^2} ]}
+ \int\limits_{ROI_z} \dd z \Psi^{ROI_{xy}}_{FE}(z)
 													  \qty[\var{D_{FP,dir}}(\var{\vb{\alpha}})
														+ \var{D_{FP,indir}}(\var{\varphi_{FP}})]
														\nonumber \\
&= \int\limits_{ROI_z} \dd z D_{FP}(z)
 \var{\overline{\xi^2}}_1(\var{N_\mathbb{A}})
 \qty[- \frac{I_{f,1}(z)}{\overline{\xi^2}(z)}
		 	+ \frac{I_{f,2}(z)}{\overline{\xi^2}(z)^2} ]
	\rightarrow \text{ direct change due to $\var{\vb*{\alpha}}$} \nonumber \\
&+ \int\limits_{ROI_z}\dd z D_{FP}(z)
   \var{\overline{\xi^2}}_2(\var{\varphi_{FP}})
		\qty[- \frac{I_{f,1}(z)}{\overline{\xi^2}(z)}
				 + \frac{I_{f,2}(z)}{\overline{\xi^2}(z)^2} ]
   \rightarrow \text{ indirect change due to $\var{\varphi_{FP}}$} \nonumber \\
&+ \int\limits_{ROI}\dd z \Psi_{FE}^{ROI_{xy}}(z)
 		\var{D_{FP,dir}}(\var{\vb*{\alpha}}, \varphi_{FP})
		\rightarrow \text{ direct change due to $\var{\vb*{\alpha}}$} \nonumber \\
&+ \int\limits_{ROI}\dd z \Psi_{FE}^{ROI}(z)
		\var{D_{FP,indir}}(\vb*{\alpha}, \var{\varphi_{FP}})
 \rightarrow \text{indirect change due to $\var{\varphi_{FP}}$}
 \label{eq:del_R_explicit}.
\end{align}

\subsection{Relating $\var{\varphi}$ to $\var{\vb*{\alpha}}$}
\label{sec:FSAP}
The expression in Equation \ref{eq:del_R_explicit} shows that the response
change depends on the changes in the system parameter vector
$\var{\vb*{\alpha}}$
and the corresponding change in the Fokker-Planck flux $\var{\varphi_{FP}}$.
Since the Fokker-Planck flux and the vector of system parameters are related
through Equation \ref{eq:1DFP}
and its associated boundary conditions it must also be that the perturbations
in both of these quantities are related. A first-order relationship between $
\var{\varphi_{FP}}$ and $ \var{\vb*{\alpha}}$ can be obtained by taking the
Gateaux-differential of the Fokker-Planck equation and it's
boundary conditions \cite{cacuciSensitivityUncertaintyAnalysis2003a}.
In this process a new PDE is obtained for the unknown
$ \var{\varphi_{FP}}$ as a function of the initial operator
$ L(\vb*{\alpha}) $ and the perturbations in the system parameters
$ \var{\vb*{\alpha}}$. Given that this PDE has to be solved for each new vector
of system parameter perturbations and the number of such vectors is large,
this process is computationally too expensive to implement in practice.

\subsection{Adjoint sensitivity analysis procedure}

An alternative to the procedure from subsection \ref{sec:FSAP} is the
Adjoint Sensitivity Analysis Procedure (ASAP).
ASAP aims to eliminate the unknown value of $\var{\varphi_{FP}}$ from the
response change Equation \ref{eq:del_R_explicit}.
This is done by constructing a new system called
the adjoint system that is independent of $\var{\varphi_{FP}}$ with the
property that the solution of this system (denoted by $\varphi^\dagger$)
can be used to eliminate the unknown
$\var{\varphi_{FP}}$ from Equation \ref{eq:del_R_explicit}.
In the process of constructing the adjoint system the boundary conditions
that ensure its unique solution will also have to be imposed. These must be
chosen such that
\cite{cacuciSensitivityUncertaintyAnalysis2003a}:
\begin{itemize}
	\item they are independent of $\var{\varphi_{FP}}$, $\var{\vb*{\alpha}}$ and
	      Gateaux-derivatives with respect to $\vb*{\alpha}$, and
	\item the evaluation of boundary terms does not contain unknown
				values of $\var{\varphi_{FP}}$.
\end{itemize}

\subsection{Adjoint system derivation}
The starting point of the adjoint system derivation is the inner product
between the adjoint flux and the operator acting on the perturbation
$\var{\varphi_{FP}}$, namely
\begin{align}
	\langle \varphi^\dagger, L(\vb*{\alpha}) \var{\varphi_{FP}} \rangle &=
	\int\limits_0^\infty \dd z \int\limits_0^\infty \dd E
	\varphi^\dagger \qty[\pdv{ \var{\varphi_{FP}}}{z}
										 - \pdv{S  \var{\varphi_{FP}}}{E}
										 - \pdv{E} \qty(T  \pdv{\var{\varphi_{FP}}}{E})
										 + \Sigma_a  \var{\varphi_{FP}}]
	\label{eq:ip_phi_adj_L_var_phi}
\end{align}
At this point we extend $\varphi_{FP}$ and consequently $\var{\varphi_{FP}}$
to the whole $\mathbb{R}^2$ plane with the condition that these quantities
are zero everywhere outside of the computational domain
$\mathscr{D}$. Through partial integration, Equation
\ref{eq:ip_phi_adj_L_var_phi} is found to be equal to
\begin{align}
	\langle \varphi^\dagger, L(\vb*{\alpha}) \var{\varphi_{FP}} \rangle
	&= \int\limits_0^\infty \dd E \varphi^\dagger(0,E) \var{\varphi_{FP}}(0,E)
	+ \left\langle - \pdv{ \varphi^\dagger}{z}
	+ S \pdv{\varphi^\dagger}{E}
	- \pdv{E}  T  \pdv{ \varphi^\dagger}{E}
	+ \Sigma_a \varphi^\dagger, \var{\varphi_{FP}}  \right\rangle \nonumber \\
	&= \int\limits_0^\infty \dd E \varphi^\dagger(0,E) \var{\varphi_{FP}}(0,E)
	+ \left\langle L^\dagger(\alpha^0) \varphi^\dagger, \var{\varphi_{FP}}
	 	\right\rangle.
	\label{eq:ip_phi_adj_L_var_phi_computed}
\end{align}
In the process of deriving Equation \ref{eq:ip_phi_adj_L_var_phi_computed}
the adjoint operator $L^\dagger$ together with its associated boundary
conditions were found to be
\begin{align}
	L^\dagger \varphi^\dagger &=
	 - \pdv{ \varphi^\dagger}{z} + S \pdv{\varphi^\dagger}{E}
	 - \pdv{E} \qty(T  \pdv{\varphi^\dagger}{E})
	 + \Sigma_a \varphi^\dagger \\
	&\text{BCE: } \varphi^\dagger(z,E_{min}) =0,
								\pdv{\varphi^\dagger}{E}\eval{}_{E=E_{min}} = 0, \\
	&\text{BCS: } \varphi^\dagger(z_{max},E) = 0.
\end{align}
To achieve the desired $\var{\varphi_{FP}}$ elimination from Equation
\ref{eq:del_R_explicit} we note that $\var{R}$ is linear in both
$\var{\vb*{\alpha}}$ and $\var{\varphi_{FP}}$. This allows writing the
Gateaux-differential of the response as
\cite{cacuciSensitivityUncertaintyAnalysis2003a}
\begin{align}
	\var{R(\vb*{e}^0; \vb*{h})} &= R'_\varphi(\vb*{e}^0) \var{\varphi_{FP}} + R'_\alpha(\vb*{e}^0) \var{\vb*{\alpha}}.
\end{align}
The quantity denoted as $R'_\varphi(\vb*{e}^0) \var{\varphi_{FP}}$,
identified as
\begin{align*}
	R'_\varphi(\vb*{e}^0) \var{\varphi_{FP}} =
	\int\limits_{ROI_z}\dd z D_{FP}(z)
	\var{\overline{\xi^2}}_2(\var{\varphi_{FP}})
	\qty[- \frac{I_{f,1}(z)}{\overline{\xi^2}(z)}
	     + \frac{I_{f,2}(z)}{\overline{\xi^2}(z)^2} ]
 + \int\limits_{ROI}\dd z \Psi_{FE}^{ROI_{xy}}(z)
  													\var{D_{FP,indir}}(\var{\varphi_{FP}}),
\end{align*}
is itself also linear in $ \var{\varphi_{FP}}$.
Coupling this with the self-duality of
Hilbert spaces certifies the application of Riesz's representation theorem.
Using this theorem the quantity $R'_\varphi(\vb*{e}^0) \var{\varphi_{FP}}$ can
written as an inner product between a quantity
$r^\dagger \in \mathscr{H}$ and $\var{\varphi_{FP}}$, namely
\cite{cacuciSensitivityUncertaintyAnalysis2003a},
\begin{align}
	R'_\varphi(\vb*{e}^0) \var{\varphi_{FP}}
	= \langle r^\dagger, \var{\varphi_{FP}} \rangle.
\end{align}
Identifying $r^\dagger$ as the right-hand
side of the adjoint system allows writing the G\^{a}teaux-differential
of the response as
\begin{align}
\var{R(\vb*{e}^0; \vb*{h})}
	&= \langle r^\dagger, \var{\varphi_{FP}} \rangle + R'_\alpha(\vb*{e}^0) \var{\vb*{\alpha}}
	= \langle L^\dagger(\vb*{\alpha}) \varphi^\dagger, \var{\varphi_{FP}} \rangle
	+ R'_\alpha(\vb*{e}^0) \var{\vb*{\alpha}}.
	\end{align}
The inner product in the second equality has already been computed in
Equation \ref{eq:ip_phi_adj_L_var_phi_computed}. Thus,
\begin{align}
	\var{R(\vb*{e}^0; \vb*{h})}	&=
	\langle \varphi^\dagger, L(\vb*{\alpha})\var{\varphi_{FP}} \rangle
	-  \int\limits_0^\infty \dd E \varphi^\dagger(0,E) \var{\varphi_{FP}}(0,E)
  + R'_\alpha(\vb*{e}^0) \var{\vb*{\alpha}}.
\end{align}
The quantity $ L(\vb*{\alpha})\var{\varphi_{FP}}$ can be derived
by taking the Gateaux-differential of the Fokker-Planck equation, and is
found to be
\begin{align}
	L(\vb*{\alpha}) \var{\varphi_{FP}}
	&= - \qty[L'_\alpha(\vb*{\alpha}) \varphi_{FP}]
	\var{\vb*{\alpha}} =
	 \pdv{\var{S} \varphi_{FP}}{E}
	+ \pdv{E} \qty(\var{T}\pdv{\varphi_{FP}}{E})
	- \var{\Sigma_a} \varphi_{FP}.
\end{align}
Making
use of this transforms $\var{R(\vb*{e}^0; \vb*{h})}$ to
\begin{align}
	\var{R(\vb*{e}^0; \vb*{h})} &=
	\left\langle \varphi^\dagger,
	- \qty[L'_\alpha(\alpha) \varphi_{FP}] \var{\vb*{\alpha}} \right\rangle
	-  \int\limits_0^\infty \dd E \varphi^\dagger(0,E) \var{\varphi}(0,E)
	+ R'_\alpha(\vb*{e}^0) \var{\vb*{\alpha}},
	\label{eq:del_R_transformed}
\end{align}
where the first inner product is equal to
\begin{align}
	\left\langle
	\varphi^\dagger, - \qty[L'_\alpha(\alpha) \varphi_{FP}] \var{\vb*{\alpha}}
	\right\rangle
	= \int_0^\infty \dd z \int_0^\infty \dd E \varphi^\dagger
	\qty[\pdv{E} \var{S} \varphi_{FP}
	+ \pdv{E} \var{T}\pdv{\varphi_{FP}}{E}
	- \var{\Sigma_a} \varphi_{FP}].
	\label{eq:indirect_effect}
\end{align}
For simplicity, the quantity $\var{\varphi}(0,E)$ was set to zero in the case
of this work and $ R'_\alpha(\vb*{e}^0) \var{\vb*{\alpha}} $ is the
direct change that has been previously computed in Equation
\ref{eq:del_R_explicit} as
\begin{align}
	R'_\alpha(\vb*{e}^0) \var{\vb*{\alpha}} =
	\int\limits_{ROI_z} \dd z D_{FP}(z)
	\var{\overline{\xi^2}}_1(\var{N_i})
	\qty[- \frac{I_{f,1}(z)}{\overline{\xi^2}(z)}
	 		 + \frac{I_{f,2}(z)}{\overline{\xi^2}(z)^2} ]
   +	\int\limits_{ROI_z}\dd z \Psi_{FE}^{ROI}(z)
		  \var{D_{FP,dir}}(\var{\vb{\alpha}}).
	\label{eq:direct_effect}
\end{align}
As can be seen from
\Crefrange{eq:del_R_transformed}{eq:direct_effect}
the goal of the ASAP has been reached. The indirect
change in the response due to the perturbation of the
Fokker-Planck flux $\var{\varphi_{FP}}$ has been replaced in Equation
\ref{eq:del_R_explicit} by the inner product from Equation
\ref{eq:indirect_effect}. Thus, for a given number $N$ of different
vectors of system parameters $\vb*{\alpha}$ the computational expense has been
decreased from the initial $N$ necessary solutions of the 1DFP system
to just two solutions, namely those of the adjoint and Fokker-Planck systems,
with a similar computational expense for both systems.

%% file: src/results.tex
\section{Results and discussion}
\label{sec:results_discussion}

This section details the computational set-up of the algorithm in Subsection
\ref{subsec:computational_setup}, the comparison between the dose calculation
of our engine and those of TOPAS and Bortfeld's method in Subsection
\ref{subsec:forward_results} and the accuracy of the previously illustrated ASAP methodology for response change computations in Subsection
\ref{subsec:adjoint_results}.

\subsection{Computational set-up}
\label{subsec:computational_setup}

\begin{figure}[!h]
\centering
  \includegraphics[width=0.6\linewidth]{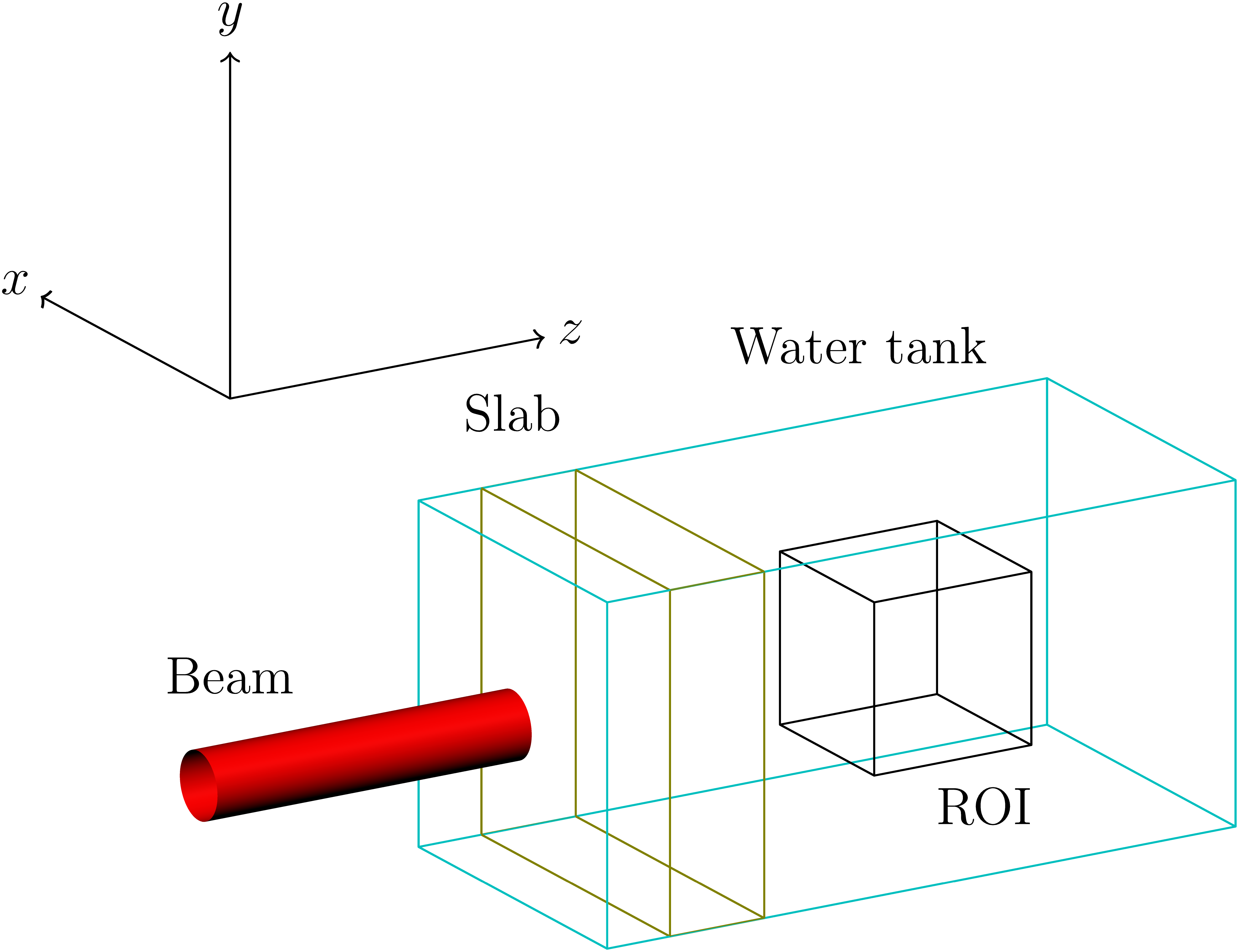}
  \caption{Illustration of the CT scan, the slab with perturbed HU values, the ROI and the incident beam}
  \label{fig:geometry_setup}
\end{figure}

The domain of the CT scan was discretized using an arbitrarily chosen number
of 51 bins in the X and Y direction and 100 bins in the Z direction. The
spatial extent of the CT scan was set to -2 to 2 cm in the X and Y
directions and 0 to 10 cm in the Z direction. Within this geometry a slab
was placed along the depth of the tank with its depth and precise position
being variable. The slab had variable HU values set while the rest of the
tank was set to either 0 HU (water) or the arbitrarily chosen value of 550 HU.
The ROI was defined to be a box with variable extents in all
three directions. An illustration of this set-up can be seen in Figure
\ref{fig:geometry_setup}

In all test cases the beam started at the point of
$\vb*{r}_{start} = (0, 0, 0)$ and ended at $\vb*{r}_{end} = (0, 0, 10)$.
The tracking of the beam within the geometry was performed using an in-house
ray-tracing procedure based on work of de Sutter
et al. \cite{sundermannFastAlgorithmCalculate1998}. A von Neumann stability
analysis was not performed however, through empirical observations
it was found that in water the maximal step size for accurate, artifact free
fluxes outputted by the CN scheme is 0.01 cm. This was used in a wrapper
function for the ray-tracing procedure to divide each segment into
a corresponding number of sub-segments.

The initial spread of the
Gaussian $\sigma_\xi$ from Equation \ref{eq:FE_init_coeffs}
in X and Y was set to 0.3 cm and the space angle correlation
$\varrho$ from Equation \ref{eq:FE_init_coeffs} was set to zero. Due to the
singularity in the angular spread variable of the coefficients from
Equation \ref{eq:FE_init_coeffs}, $\sigma_\omega$ could not be set to zero.
However, it was found that values below $10^{-4}$ did not
affect the resulting energy deposition distributions and thus the angular
spread was set to the dimensionless value of $10^{-8}$. The energy domain was
fixed between $E_{min} = \SI{1}{\MeV}, E_{max} = \SI{100}{\MeV}$ with
a number of groups of $NG = 300$. The coefficients of the energy initial
condition from Equation \ref{eq:1DFP_BCE} were set to correspond to a normal
distribution and were matched such that the number of particles was either 1 or
$\num{2 E7}$. Moreover, the initial beam average energy $E_0$ from Equation
\ref{eq:1DFP_BCS} was set to $\SI{100}{\MeV}$ while the energy spread
$\sigma_E$ from the same equation was set to $\SI{0.757504}{\MeV}$. The
energy spread value was chosen to match the standard value that TOPAS
initializes for a proton pencil beam.

\subsection{Forward results}
\label{subsec:forward_results}
\begin{figure}[!h]
  \vspace{-.4cm}
  \centering
  \includegraphics[width=.6\linewidth]{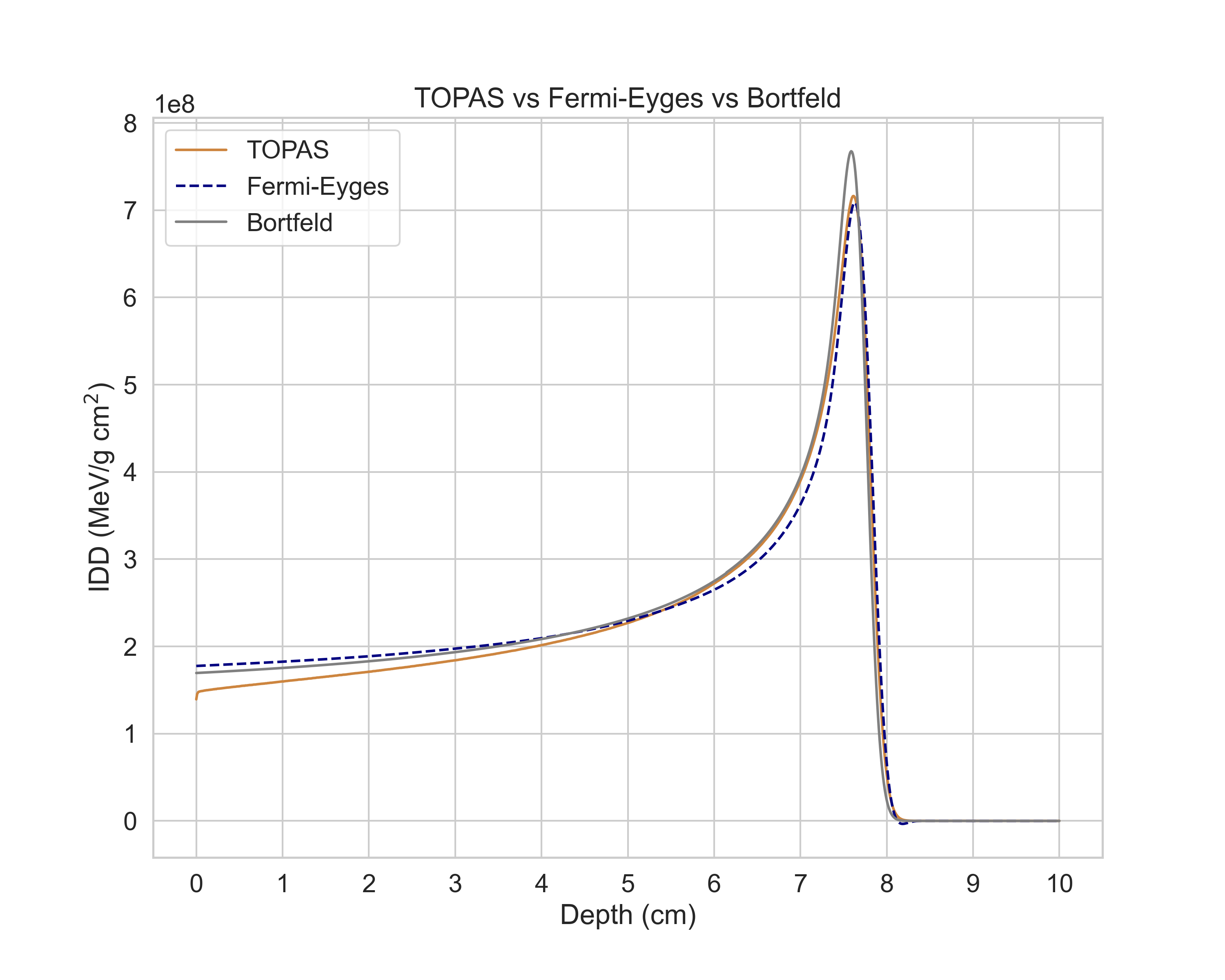}
  \caption{IDD of the in-house algorithm, TOPAS and Bortfeld's model}
  \label{fig:IDD_bortfeld}
\end{figure}

To gauge the accuracy of the reponse computation engine, it was benchmarked
against the TOPAS MC algorithm \cite{perlTOPASInnovativeProton2012a}. In a
homogeneous water tank the energy deposition can readily be converted to
dose deposition. Laterally integrating this three dimensional quantity
results in the dose-depth curve named integrated depth dose (IDD). The
comparison against TOPAS and Bortfeld's popular pencil beam algorithm
\cite{bortfeldAnalyticalApproximationBragg1997a}
can be seen in Figure \ref{fig:IDD_bortfeld} showing that
our algorithm is capable of accurately predicting the dose in the
Bragg peak, the region of most clinical interest. Our algorithm slightly
over-estimates the dose in the entrance region, due
to the assumption that $\SI{100}{\percent}$ of the energy released in nuclear
interactions is deposited locally. A refinement of the treatment of nuclear
interactions can be envisioned using convolutional methods or through more
simplistic fits to empirical data. However, this is not the purpose of this
paper, whose main focus is the sensitivity of the algorithm.

\subsection{Adjoint results}
\label{subsec:adjoint_results}

\input{./src/tables/adjoint_cases_results_overview.tex}

A variety of test cases for the response change computation was
performed, an overview of which can be seen in Table
\ref{tab:adjoint_test_cases_overview}.
For all test cases $\Sigma_a$ was set to zero due to a lack of
data for all other materials other than 0 HU.

\subsubsection{Case I}
\begin{figure}[!h]
  \vspace{-.35cm}
  \centering
  \includegraphics[width=.5\linewidth]{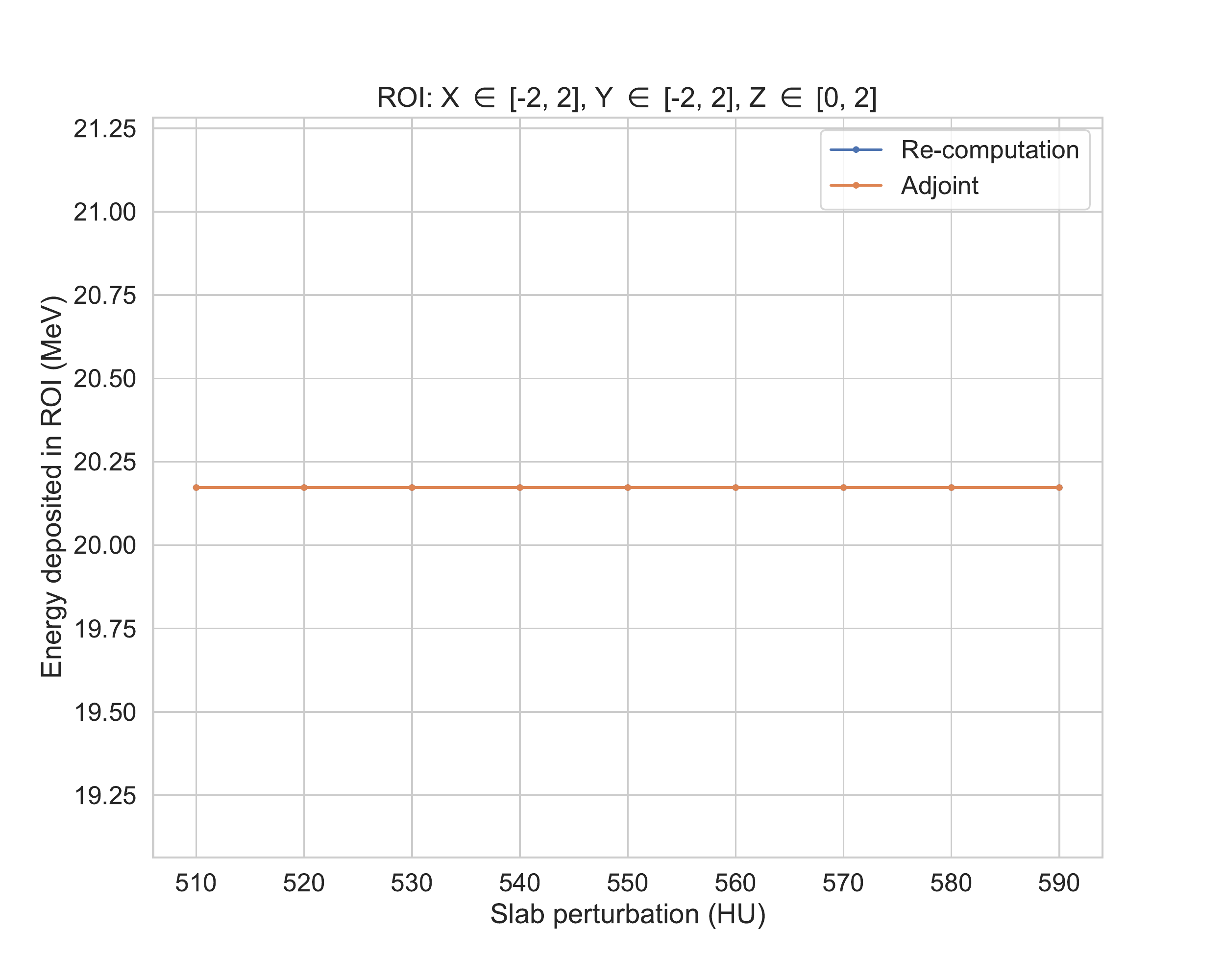}
  \caption{Adjoint versus re-computation for a small HU range with a full lateral ROI in X and Y and Z $\in [0, 2]$.}
  \label{fig:adjoint_results_corectness_test}
\end{figure}

First, a small range of [-40, 40] HU perturbations around the nominal slab
value of 550 HU was used. The slab was placed in the plateau region of the
energy deposition versus depth curve between 2 and 3 cm in depth.
Given this set-up if the ROI has dimensions $X, Y \in [-2, 2], Z \in [0, 2] $
and under the assumption of a forward propagating (not backscattered) flux
then the expectation is that the response change is identically zero.
This corectness test can be seen in Figure
\ref{fig:adjoint_results_corectness_test}.

\subsubsection{Case II}

\begin{figure}[!h]
  \vspace{-.35cm}
  \centering
  \includegraphics[width=.5\linewidth]{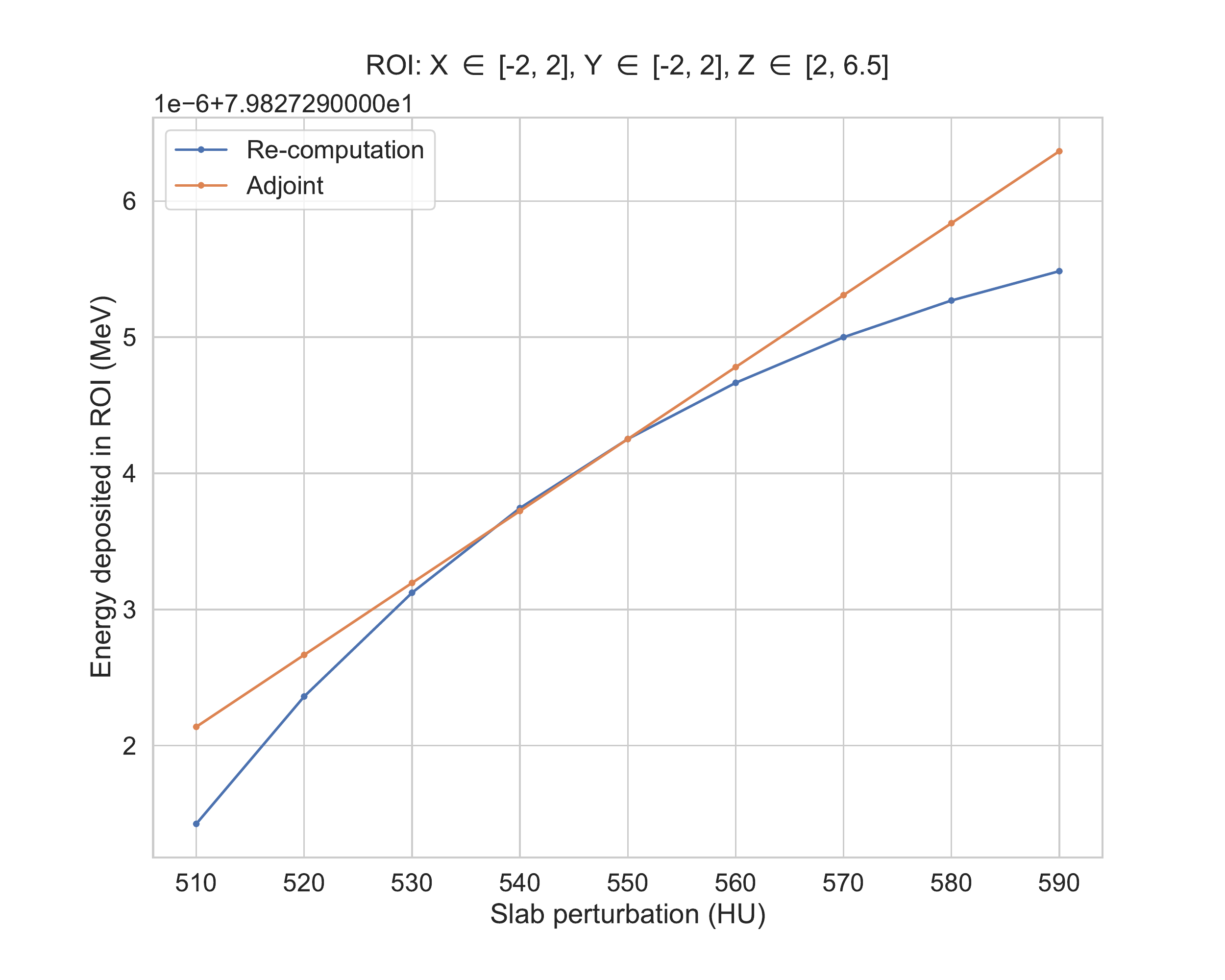}
  \caption{Adjoint versus re-computation for a small HU range with a full lateral ROI in X and Y, Z $\in [2, 6.5]$.}
  \label{fig:Slab_2-3_ROI_XY_-2_2_Z_2_6.5_base_550_step=10}
\end{figure}

Changing the ROI to be the box with $X,Y \in [-2, 2], Z \in [2, 6.5]$
results in Figure \ref{fig:Slab_2-3_ROI_XY_-2_2_Z_2_6.5_base_550_step=10}.
There it can be seen that adjoint theory provides a first order approximation
to the forward response. In this case the maximal percent error occurs for the
510 HU slab perturbation and it is equal to $\approx \SI{1.1E-6}{\percent}$.

\subsubsection{Case III}

\begin{figure}[!h]
  \vspace{-.4cm}
  \centering
  \begin{minipage}{0.49\textwidth}
    \centering
    \includegraphics[width=\linewidth]{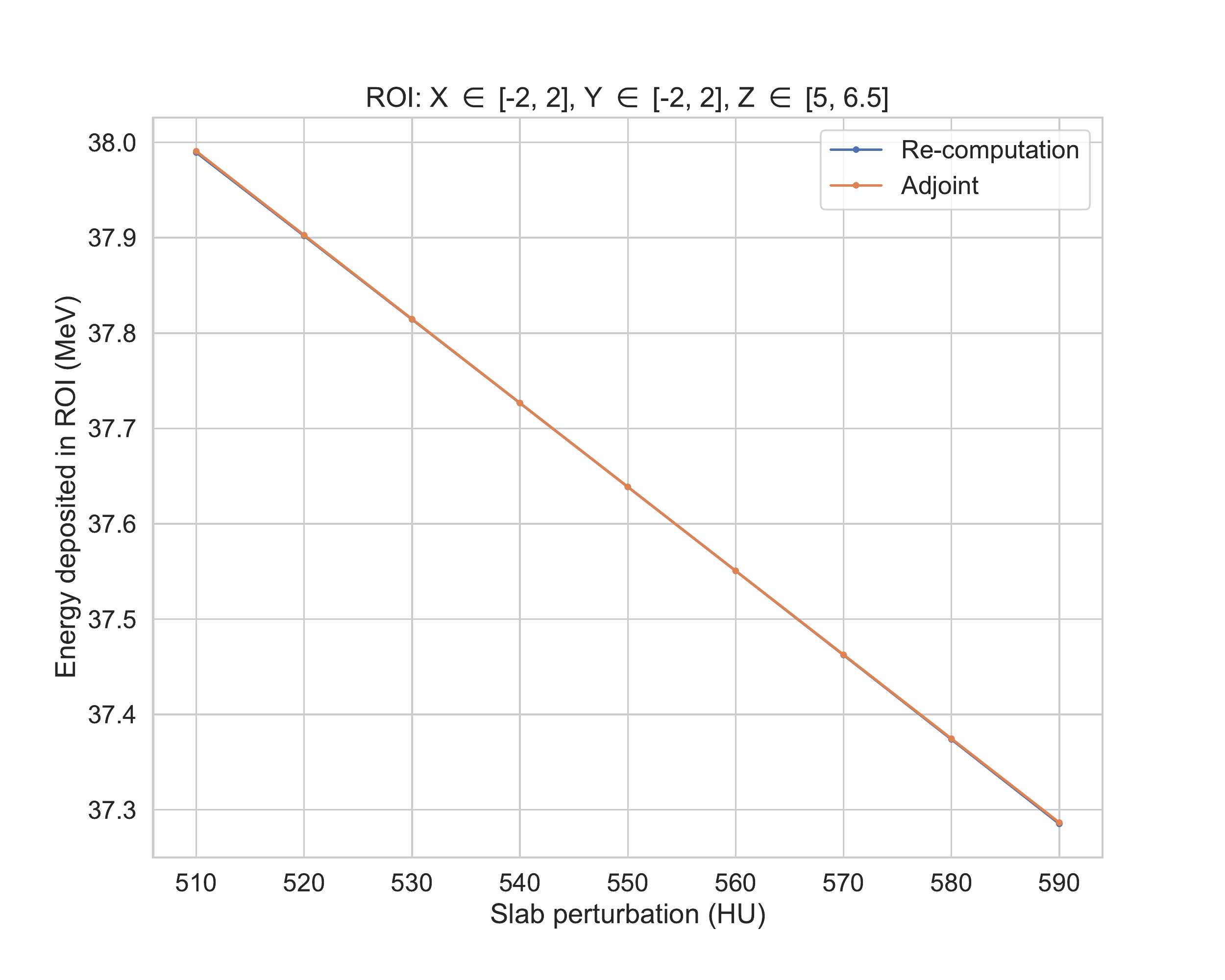}
    \caption{Adjoint versus re-computation for a small HU range with a full lateral ROI in X and Y and Z $\in [5, 6.5]$.}
    \label{fig:Slab_2-3_ROI_XY_-2_2_Z_5_6.5_base_550_step=10}
  \end{minipage}\hfill
  \begin{minipage}{0.49\textwidth}
    \centering
    \includegraphics[width=\linewidth]{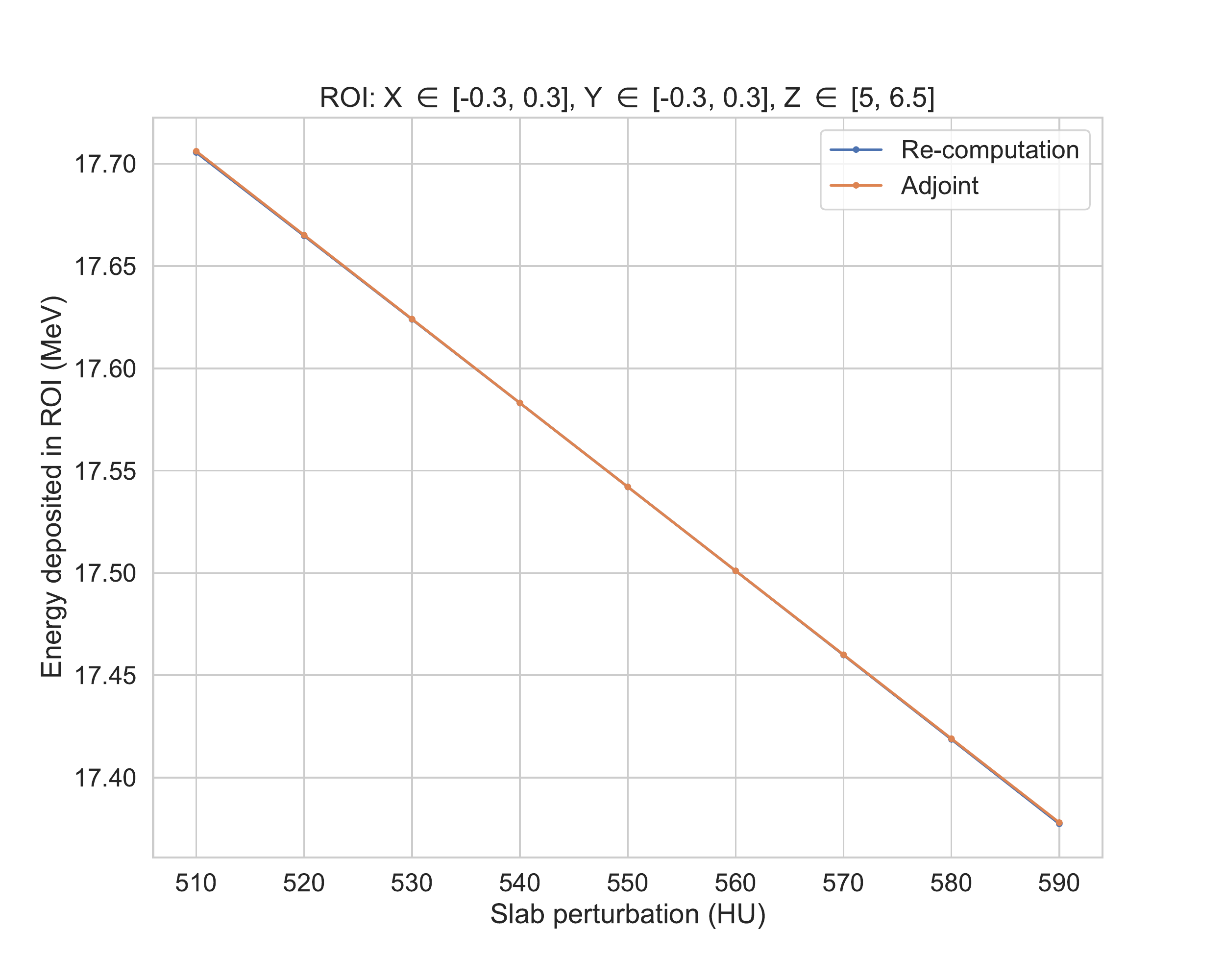}
    \caption{Adjoint versus re-computation for a small HU range with a
    reduced lateral ROI in X and Y and Z $\in [5, 6.5]$.}
    \label{fig:Slab_2-3_ROI_XY_-0.3_0.3_Z_5_6.5_base_550_step=10}
  \end{minipage}
\end{figure}

In the case of proton therapy, it is likely that a ROI of practical interest
is the Bragg Peak region. Thus, the slab is maintained in its previous position
and the ROI is set to the box with $X,Y \in [-2, 2], Z \in [5, 6.5]$. This
result can be seen in
Figure \ref{fig:Slab_2-3_ROI_XY_-2_2_Z_5_6.5_base_550_step=10}.
Another scenario of interest could be the one in which a tumor is surrounded
by organs at risk. In this case, the ROI is restricted to only part of the
lateral extent. In Figure
\ref{fig:Slab_2-3_ROI_XY_-0.3_0.3_Z_5_6.5_base_550_step=10} the lateral
extent was constrained to $X, Y \in [-0.3, 0.3]$ while the depth was kept to
$Z \in [5, 6.5]$.
Both Figure \ref{fig:Slab_2-3_ROI_XY_-2_2_Z_5_6.5_base_550_step=10}
and \ref{fig:Slab_2-3_ROI_XY_-0.3_0.3_Z_5_6.5_base_550_step=10} show that
the adjoint method is capable of accurately computing the response changes
due to the slab in the ROI down to a percentage error of
$\SI{3.6E-3}{\percent}$ that is likely clinically
insignificant. This can be seen from the fact that a fraction that is
typically delivered is on the order of $\SI{2}{\gray}$ which is equivalent to
$\approx \SI{6.3E9}{\MeV \per \gram}$.

\subsubsection{Case IV}

\begin{figure}[!h]
  \vspace{-.35cm}
  \centering
  \includegraphics[width=.5\linewidth]{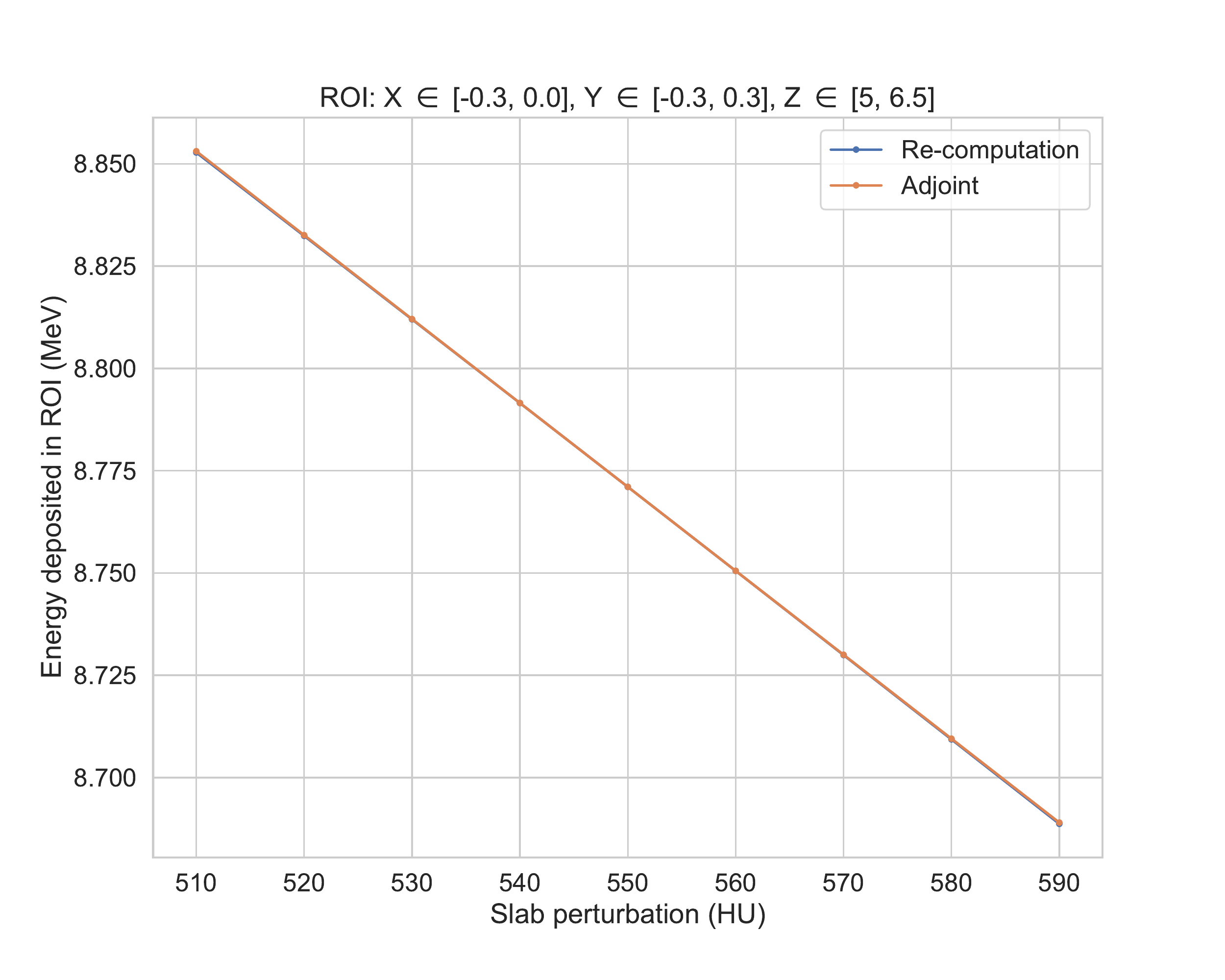}
  \caption{Adjoint versus re-computation for a small HU range with an laterally asymmetric ROI in X and Y and Z $\in [5, 6.5]$.}
  \label{fig:Slab_2-3_ROI_X_-0.3_0_Y_-0.3_0.3_Z_5_6.5_base_550_step=10}
\end{figure}

Tests with asymmetric ROIs have also been performed.
Restricting the ROI such that $X \in [-0.3,0]$, $Y \in [-0.3, 0.3]$ and
keeping $Z \in [5, 6.5]$ resulted in a similar error (depicted as the same
in Table \ref{tab:adjoint_test_cases_overview} due to round-off) to the
previous test cases. The two curves for this case can be seen in Figure
\ref{fig:Slab_2-3_ROI_X_-0.3_0_Y_-0.3_0.3_Z_5_6.5_base_550_step=10}.

\subsubsection{Case V}

\begin{figure}[!h]
  \centering
  \includegraphics[width=.6\linewidth]{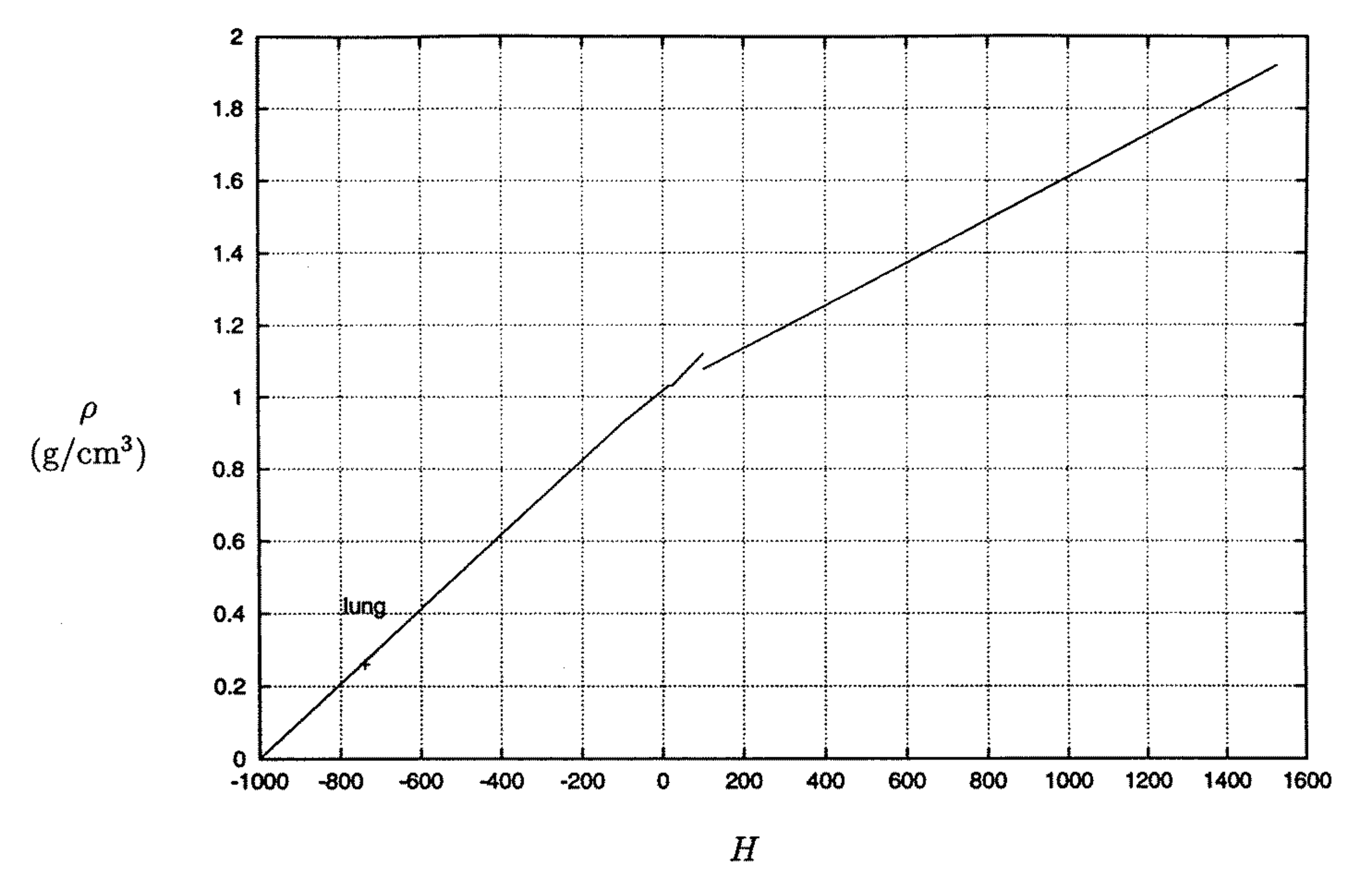}
  \caption{Illustration of the discontinuity in Schneider's HU to density conversion curve. Picture taken from \cite{schneiderCorrelationCTNumbers2000}}
  \label{fig:Schneider_density_to_HU}
\end{figure}

\begin{figure}[!h]
  \vspace{-.4cm}
  \centering
  \begin{minipage}{0.49\textwidth}
    \centering
    \includegraphics[width=\linewidth]{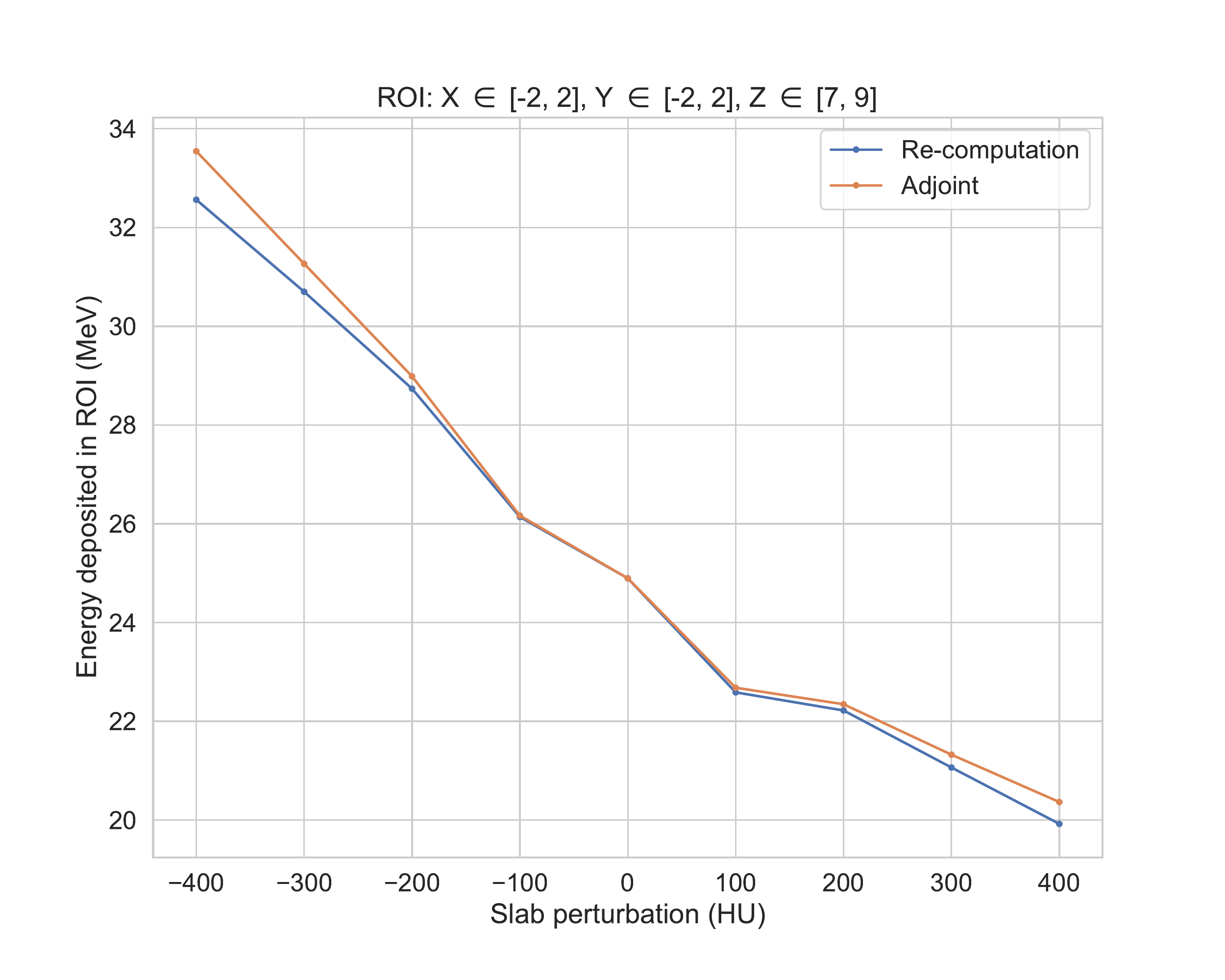}
    \caption{Adjoint versus re-computation for a large HU range with a laterally symmetric ROI and Z $\in [7, 9]$.}
    \label{fig:Slab_2-3_ROI_XY_-2_2_Z_7_9_base_0_step=100}
  \end{minipage}\hfill
  \begin{minipage}{0.49\textwidth}
    \centering
    \includegraphics[width=\linewidth]{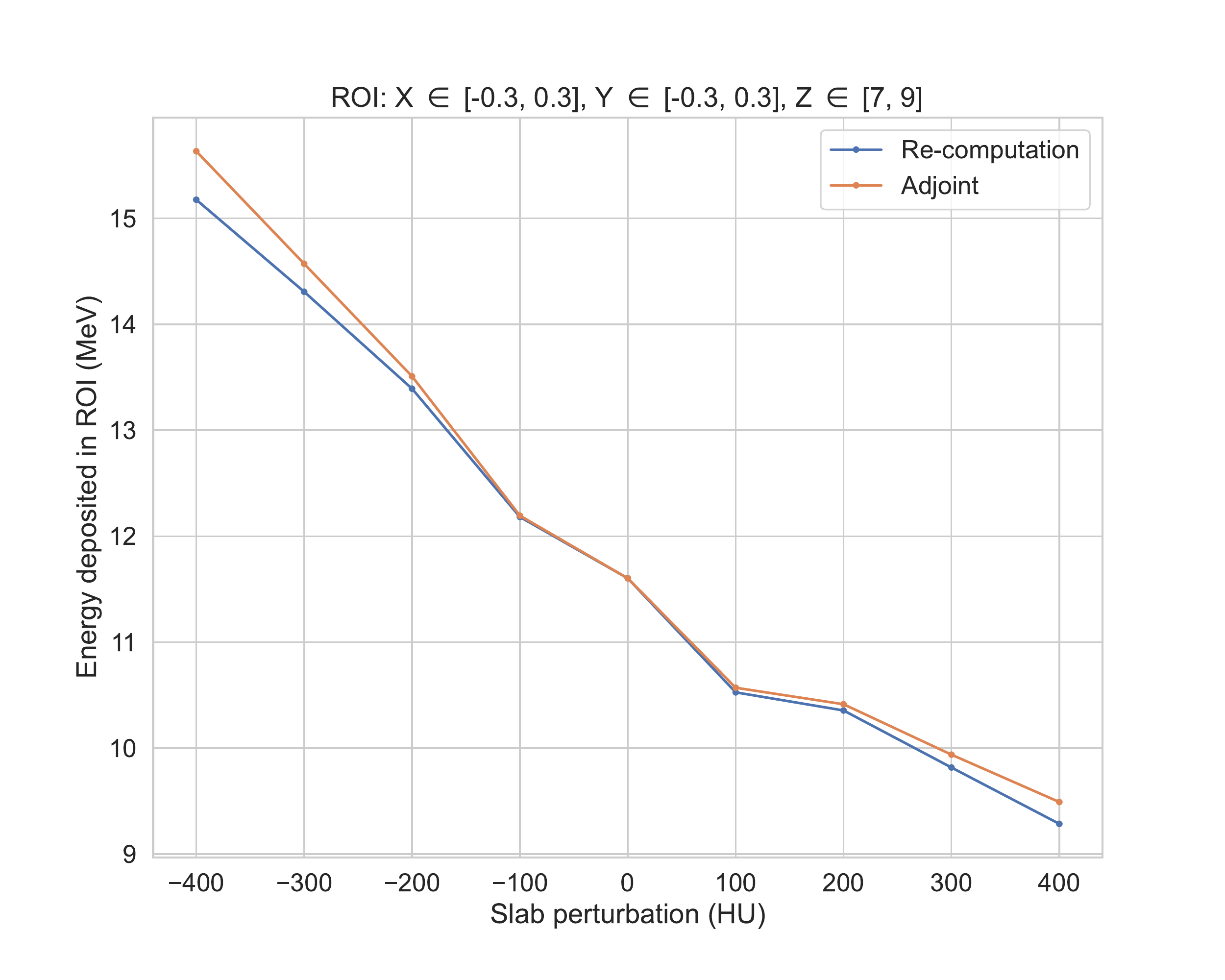}
    \caption{Adjoint versus re-computation for a large HU range with a laterally symmetric reduced ROI and Z $\in [7, 9]$.}
    \label{fig:Slab_2-3_ROI_XY_-0.3_0.3_Z_7_9_base_0_step=100}
  \end{minipage}
\end{figure}

Next to the small HU range, a large [-400, 400] HU perturbation range
around the nominal value of 0 HU was tested. The nominal value in this case
corresponded to a homogeneous water tank. This set-up simulates more
clinically relevant test-cases as the -400 HU value roughly corresponds
to a tissue similar to lung while a value of 400 HU correponds to bone.
As in the small perturbation range cases, ROIs with full and reduced
lateral X and Y extents were tested. Figure
\ref{fig:Slab_2-3_ROI_XY_-2_2_Z_7_9_base_0_step=100} illustrates
the case when $X \in [-2,2]$, $Y \in [-2, 2], Z \in [7, 9]$ while
Figure \ref{fig:Slab_2-3_ROI_XY_-0.3_0.3_Z_7_9_base_0_step=100}
illustrates the case when $X \in [-0.3, 0.3]$, $Y \in [-0.3, 0.3],
Z \in [7, 9]$. It should be noted that as opposed to the straight lines
shown before, these figures do not contain straight lines. This is
due to the regions of discontinuity that appear in the Scheinder's
conversion \cite{schneiderCorrelationCTNumbers2000} from HU values to
density and atomic composition. The discontinuity for the density to HU
converison can be seen in Figure \ref{fig:Schneider_density_to_HU} and it is
also why the range of $\numrange{510}{590}$ was chosen initially,
as in this range the afore mentioned conversion is continuous.

\subsubsection{Case VI}

Tests were also performed with a 2 cm slab placed between 4 and 6 cm deep,
in the vicinity of the BP. The same large and small variations in the
perturbation range together with ROI contractions were investigated. In the
case of a small perturbation range, good agreement was found for both full and
symmetrically reduced lateral X and Y extents with maximal
percentage errors being $\SI{3.6E-3}{\percent}$. In the case of the large
perturbation range, the algorithm resulted in a moderate percentage error of
$\SI{17}{\percent}$. This result is not unexpected as adjoint theory provides
a first order approximation to the response change and it is expected that the
approximation worsens as a function of increasing perturbations.

\begin{figure}[!h]
  \vspace{-.4cm}
  \centering
  \begin{minipage}{0.49\textwidth}
      \centering
      \includegraphics[width=\linewidth]{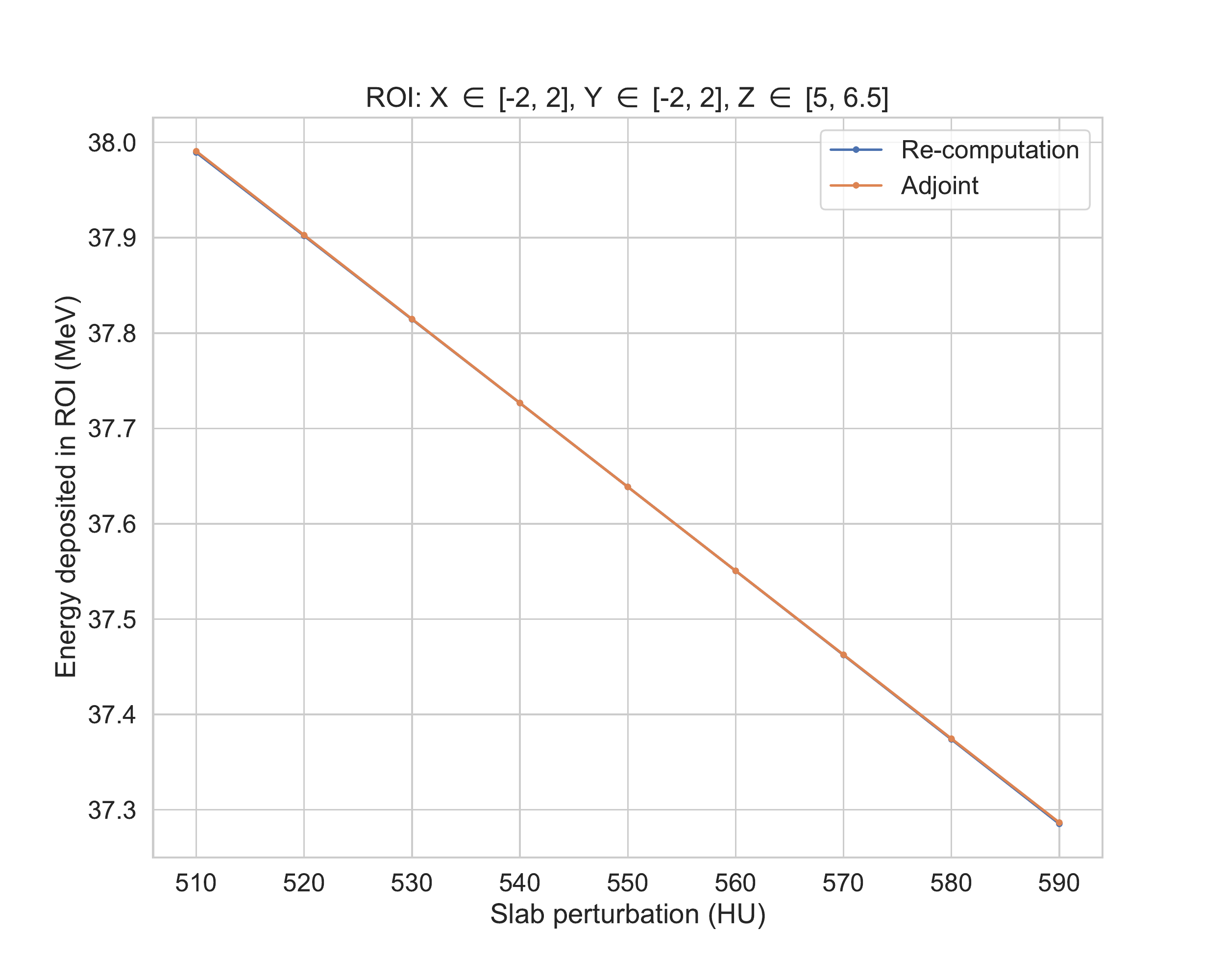}
      \caption{Adjoint versus re-computation for a small HU range with a laterally symmetric ROI and Z $\in [5, 6.5]$.}
      \label{fig:base_550_slab_4-6_ROI_XY_-2_2_Z_5_6.5}
  \end{minipage}\hfill
  \begin{minipage}{0.49\textwidth}
      \centering
      \includegraphics[width=\linewidth]{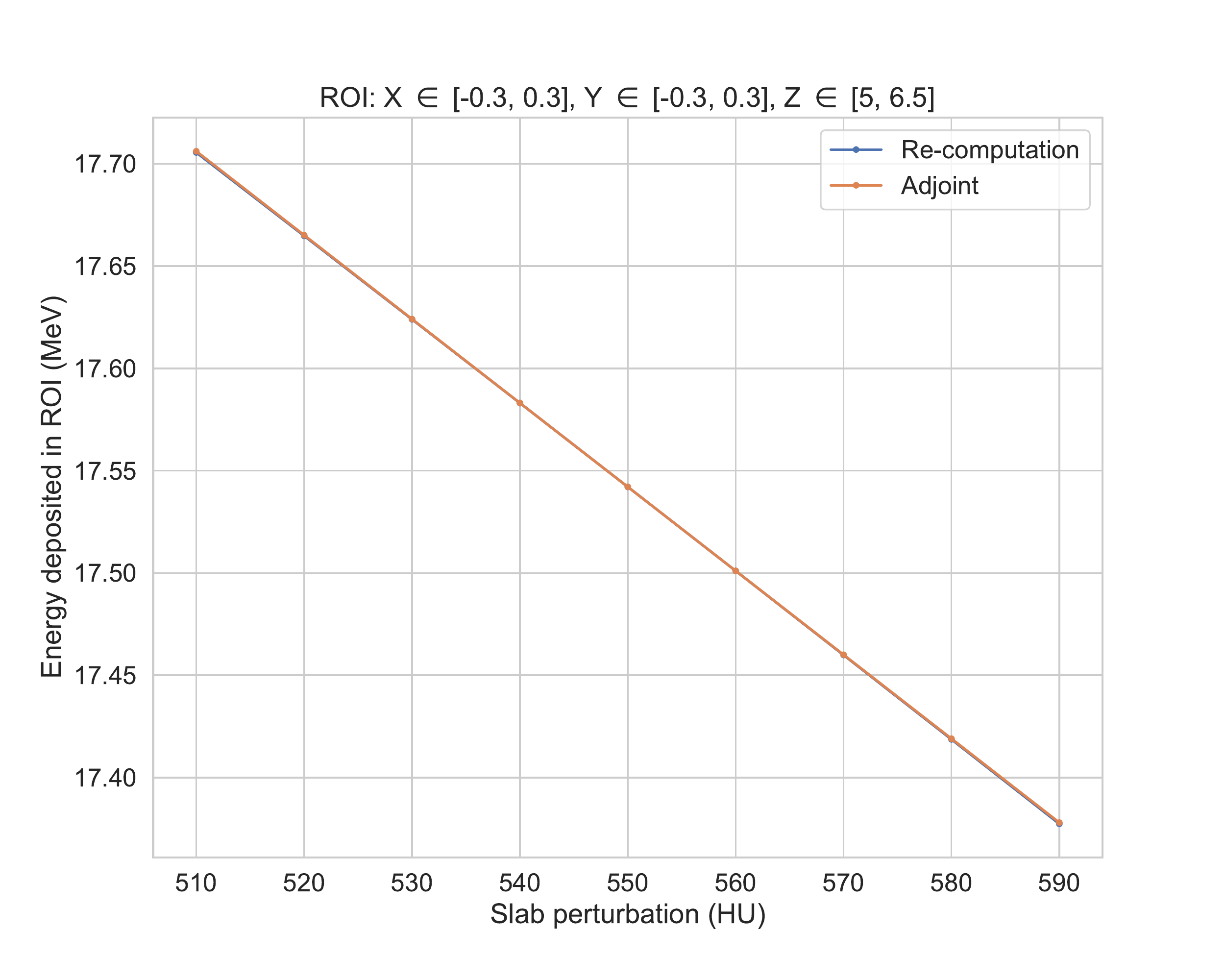}
      \caption{Adjoint versus re-computation for a large HU range with a laterally reduced symmetric ROI and Z $\in [5, 6.5]$.}
  \end{minipage}
\end{figure}

\begin{figure}[!h]
  \vspace{-.4cm}
  \centering
  \begin{minipage}{0.49\textwidth}
      \centering
      \includegraphics[width=\linewidth]{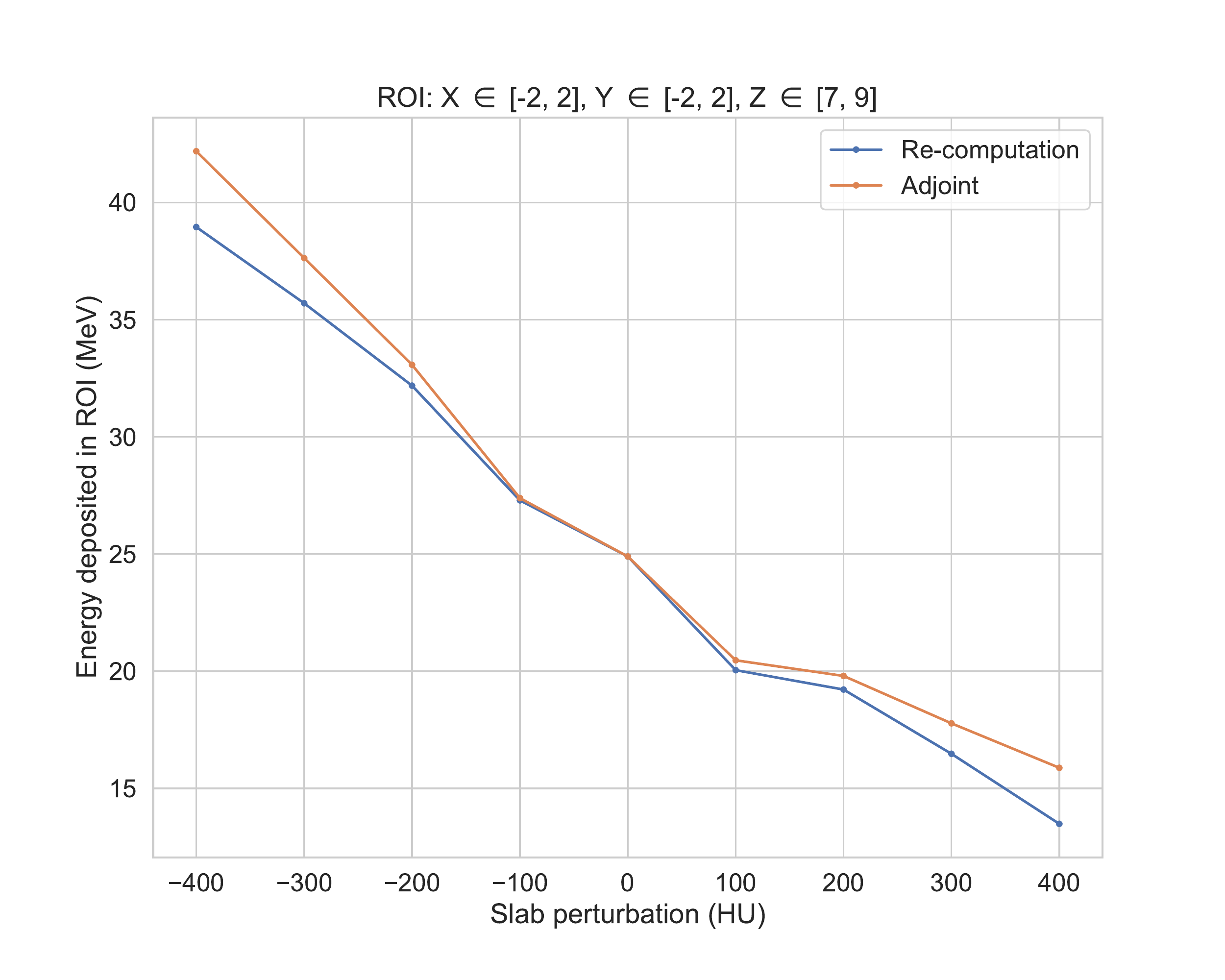} 
      \caption{Adjoint versus re-computation for a large HU range with a laterally symmetric ROI and Z $\in [7, 9]$.}
  \end{minipage}\hfill
  \begin{minipage}{0.49\textwidth}
      \centering
      \includegraphics[width=\linewidth]{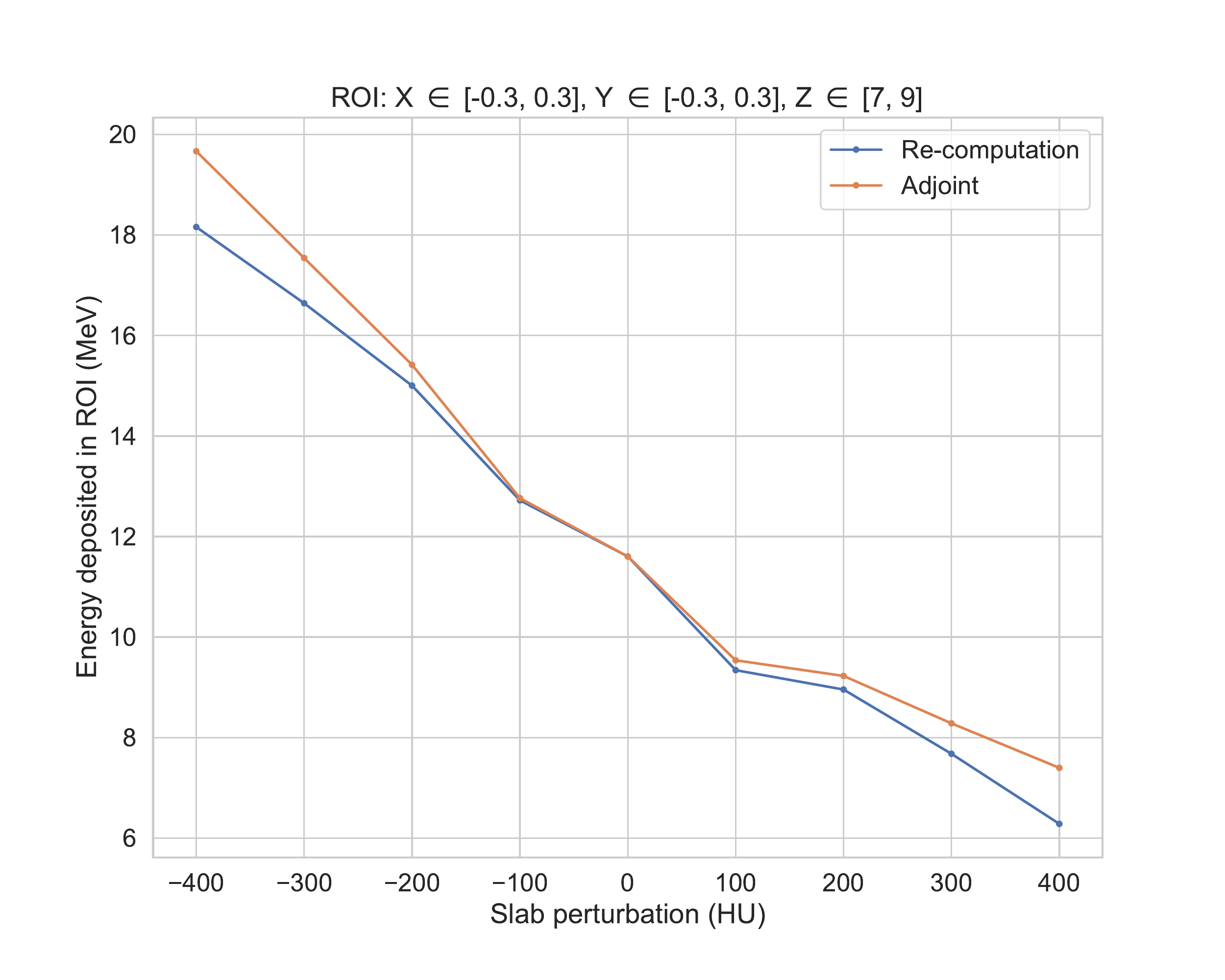} 
      \caption{Adjoint versus re-computation for a large HU range with a laterally symmetrically reduced ROI and Z $\in [7, 9]$.}
  \end{minipage}
\end{figure}

%% file: src/tables/adjoint_cases_results_overview.tex
\begin{table}[!h]
\centering
\resizebox{\textwidth}{!}{%
\begin{tabular}{|c|cc|ccc|c|}
\hline
\multirow{2}{*}{\textbf{\begin{tabular}[c]{@{}c@{}}Case \\ number\end{tabular}}} & \multicolumn{2}{c|}{\textbf{Slab}}              & \multicolumn{3}{c|}{\textbf{ROI extent}}                                       & \multirow{2}{*}{\textbf{\begin{tabular}[c]{@{}c@{}}Maximal percentage \\ error\end{tabular}}} \\
                                                                                 & Location (cm)               & Perturbation (HU) & X (cm)                           & Y (cm)                           & Z (cm)   &                                                                                               \\ \hline
I                                                                                & \multicolumn{1}{c|}{[2, 3]} & [-40, 40]         & \multicolumn{1}{c|}{[-2, 2]}     & \multicolumn{1}{c|}{[-2, 2]}     & [0, 2]   & 0                                                                                             \\
II                                                                               & \multicolumn{1}{c|}{[2, 3]} & [-40, 40]         & \multicolumn{1}{c|}{[-2, 2]}     & \multicolumn{1}{c|}{[-2, 2]}     & [2, 6.5] & $\num{1.1 E-6}$                                                                               \\
III                                                                              & \multicolumn{1}{c|}{[2, 3]} & [-40, 40]         & \multicolumn{1}{c|}{[-2, 2]}     & \multicolumn{1}{c|}{[-2, 2]}     & [5, 6.5] & $\num{3.6 E-3}$                                                                               \\
III                                                                              & \multicolumn{1}{c|}{[2, 3]} & [-40, 40]         & \multicolumn{1}{c|}{[-0.3, 0.3]} & \multicolumn{1}{c|}{[-0.3, 0.3]} & [5, 6.5] & $\num{3.6 E-3}$                                                                               \\
IV                                                                               & \multicolumn{1}{c|}{[2, 3]} & [-40, 40]         & \multicolumn{1}{c|}{[-0.3, 0]}   & \multicolumn{1}{c|}{[-0.3, 0.3]} & [5, 6.5] & $\num{3.6 E-3}$                                                                               \\
V                                                                                & \multicolumn{1}{c|}{[2, 3]} & [-400, 400]       & \multicolumn{1}{c|}{[-2, 2]}     & \multicolumn{1}{c|}{[-2, 2]}     & [7, 9]   & $\num{3.0}$                                                                                   \\
V                                                                                & \multicolumn{1}{c|}{[2, 3]} & [-400, 400]       & \multicolumn{1}{c|}{[-0.3, 0.3]} & \multicolumn{1}{c|}{[-0.3, 0.3]} & [7, 9]   & $\num{3.0}$                                                                                   \\
VI                                                                               & \multicolumn{1}{c|}{[4, 6]} & [-40, 40]         & \multicolumn{1}{c|}{[-2, 2]}     & \multicolumn{1}{c|}{[-2, 2]}     & [5, 6.5] & $\num{3.6 E-3}$                                                                               \\
VI                                                                               & \multicolumn{1}{c|}{[4, 6]} & [-40, 40]         & \multicolumn{1}{c|}{[-0.3, 0.3]} & \multicolumn{1}{c|}{[-0.3, 0.3]} & [5, 6.5] & $\num{3.6 E-3}$                                                                               \\
VI                                                                               & \multicolumn{1}{c|}{[4, 6]} & [-400, 400]       & \multicolumn{1}{c|}{[-2, 2]}     & \multicolumn{1}{c|}{[-2, 2]}     & [7, 9]   & $\num{17.7}$                                                                                  \\
VI                                                                               & \multicolumn{1}{c|}{[4, 6]} & [-400, 400]       & \multicolumn{1}{c|}{[-0.3, 0.3]} & \multicolumn{1}{c|}{[-0.3, 0.3]} & [7, 9]   & $\num{17.7}$                                                                                  \\ \hline
\end{tabular}%
}
\caption{Overview of case numbers and the corresponding slab location and perturbation values, ROI values and maximal percentage errors between the re-computation and the adjoint result.}
\label{tab:adjoint_test_cases_overview}
\end{table}

%% file: src/conclusion.tex
\section{Conclusion}
\label{sec:conclusion}
In this paper we have developed a methodology for the approximate solution of
the Linear Boltzmann Equation that takes heterogeneity in the depth direction
into account. This method requires the solution to two PDEs, namely the
one-dimensional FP equation and the FE equation. The one-dimensional FP
equation was numerically solved through a
combination of the SIPG method using quadratic energy basis
functions in energy and the CN method in space. Using the 1DFP flux
$\varphi_{FP}$ the average depth-dependent energy $E_a(z)$ of the beam is
computed. This quantity is thereafter used in the computation of the
depth-dependent FE coefficients which define the FE flux $\varphi_{FE}$.
Using the product of these two fluxes complete knowledge of the
phase-space density of protons is obtained and hereby our specific problem
of charged particle transport is solved.

Using the phase-space density of protons, the response (which was defined
by the deposited energy in an arbitrary ROI) was computed.
Good agreement, especially in the clinically significant Bragg peak region, was
obtained in a homogeneous water tank when our algorithm was benchmarked against
TOPAS (taken as the reference algorithm) and Bortfeld's popular pencil beam
algorithm.

Using functional analysis the adjoint system was derived and solved.
The changes in the response due to slabs placed along the depth of the tank
with different HU values were computed. These changes were compared against the
re-computation ones with relatively good results. Adjoint theory provided
(as expected) a first order approximation to the response change curve.
In the case of small slab perturbations the relative difference was clinically
insignificant. Even in cases of large HU perturbations,
adjoint theory resulted in relatively small to moderate errors of
\qtyrange{3}{17}{\percent}.

Future work should focus both on improving the energy/dose deposition
component (forward) and on speeding up the response change
computation (backward). To improve the forward component a better model
is needed for the inelastic nuclear interactions between the primary protons
and the irradiated tissue. Possible methods for approaching this
would be a convolution-based approach or the Monte-Carlo fit method outlined
by Soukup et al. \cite{soukupPencilBeamAlgorithm2005}.
Moreover, it is clear that
our algorithm cannot in its current state account for lateral heterogeneities
and thus, a pencil beam splitting scheme is needed. A starting point for this
would be the well-performing scheme proposed by Yang et. al
\cite{yangImprovedBeamSplitting2020}.
Another metric to improve on is speed. This is an area where our algorithm
performed relatively well, with an average execution time of
\SI{0.1}{\second} for one pencil beam. Ultimately, we aim to reduce the
execution time down to the \unit{\ms} range. This can be achieved by
implementing an adaptive energy grid such that no empty energy groups are
solved for. Moreover, the process of tracking
many beams through the CT scan is highly parallelizable due to their
independent nature.

The main drawback of the adjoint component is the long time presently needed to
compute the direct contribution to the response change due to
$\var{N_\mathbb{A}}$ versus the relatively small increase in accuracy it
yields. Tabulating the integrals involved should, similar to the forward
component, yield significant time reductions. Moreover,
the necessary data for the absorption cross section should be obtained.
Another point of improvement is the inclusion of perturbations in
the initial values of the FE coefficients or of the FP boundary condition
coefficients. Such perturbations can be derived from machine log-files (for
example the difference in MU values for a point is related to a difference in
the number of input protons and the difference in the spot positions is
related to a different entry position and angle of the beam) and are
the way in which our algorithm can be used for the purpose of patient-specific
quality assurance.

\section{Conflicts of interest and acknowledgements}
The authors wish to acknowledge that the manuscript is partly funded by Varian,
a Siemens Healthineers Company and the fruitful discussions with the
‘HollandPTC consortium – Erasmus Medical
Center, Rotterdam, Holland Proton Therapy Centre, Delft,
Leiden University Medical Center (LUMC), Leiden and Delft University of
Technology, Delft, The Netherlands’.

The authors declare that they have no known competing financial interests or
personal relationships that could have appeared to influence the work reported
in this paper. Moreover, no data was used for the research described in the
article.

\section{CRediT statement}

\textbf{Tiberiu Burlacu}: Conceptualization, methodology, software,
validation, formal analysis, data curation, investigation, writing - original draft, writing - review \& editing, visualization.\\
\textbf{Danny Lathouwers:} Conceptualization, methodology, software,
validation, resources, writing - review \& editing, supervision. \\
\textbf{Zolt\'{a}n Perk\'{o}:} Conceptualization, methodology, validation,
resources, writing - review \& editing, supervision, project administration,
funding acquisition.